
%
\message{Warning! In view of the number and sizes of indices,
a reduced version may be dangerous for your eyes!}
\input harvmac.tex
\overfullrule=0pt
\Title{SPhT/92-001}
{\vbox{
\centerline{{ Combinatorics of the Modular Group II}}
\medskip
\centerline{{The Kontsevich integrals }}}}
\vskip1cm
\centerline{C. Itzykson and J.-B. Zuber}
\smallskip
\centerline{ \it   Service de Physique Th\'eorique de
Saclay\footnote*{Laboratoire de
la Direction des Sciences de la Mati\`ere du Commissariat \`a l'Energie
Atomique}, 
F-91191 Gif-sur-Yvette cedex, France}
\vskip .3in
{\bf Abstract} We study algebraic aspects of Kontsevich integrals
as generating functions for intersection theory over moduli space
and review the derivation of Virasoro and KdV constraints.
\vskip .3in
\ifx\answ\bigans
\centerline{\bf Contents}\nobreak\medskip{\baselineskip=12pt
 \parskip=0pt\catcode`\@=11
\smallskip
\noindent {0.} {Introduction} \leaderfill{1} \par
\noindent {1.} {Intersection numbers.} \leaderfill{2} \par
\noindent {2.} {The Kontsevich integral.} \leaderfill{5} \par
\noindent \quad{2.1.} {The main theorem} \leaderfill{5} \par
\noindent \quad{2.2.}
{Expansion of $Z$ on characters and Schur functions} \leaderfill{7} \par
\noindent \quad{2.3.}
{Proof of the first part of the Theorem.} \leaderfill{13} \par
\noindent {3.} {From Grassmannians to KdV} \leaderfill{19} \par
\noindent {4.}
{Matrix Airy equation and Virasoro highest weight conditions.}
\leaderfill{29} \par
\noindent {5.} {Genus expansion.} \leaderfill{31} \par
\noindent {6.} {Singular behaviour and Painlev\'e equation.}
\leaderfill{38} \par
\noindent {7.} {Generalization to higher degree potentials}
\leaderfill{41} \par
\catcode`\@=12 \bigbreak\bigskip}\else \fi

\Date{ 1/1992\ \ Submitted to Int. J. Mod. Phys.}

%
\def\tilde{\widetilde}
\def\bar{\overline}

\def\*{\star}
\def\({\left(}		
\def\){\right)}		
\def\[{\left[}		\def\BBL{\Bigr[}
\def\]{\right]}		\def\BBR{\Bigr]}

%
%
\def\frac#1#2{{#1 \over #2}}		
\def\inv#1{{1 \over #1}}

\def\d{\partial}

\def\dd#1#2{{\partial #1 \over \partial #2}}
\def\vev#1{\langle #1 \rangle}
\def\ket#1{ | #1 \rangle}
\def\bra#1{ \langle #1 |}

\def\2pi{\hbox{$2\pi i$}}

\def\dsl{\raise.15ex\hbox{/}\kern-.57em\partial}
\def\Dsl{\,\raise.15ex\hbox{/}\mkern-.13.5mu D}
%
%
		\def\CC{{\cal C}}
\def\CD{{\cal D}}		
		
		\def\CL{{\cal L}}
\def\CM{{\cal M}}

%

%


\def\IR{\relax{\rm I\kern-.18em R}}
\font\cmss=cmss10 \font\cmsss=cmss10 at 7pt
\def\IZ{\relax\ifmmode\mathchoice
{\hbox{\cmss Z\kern-.4em Z}}{\hbox{\cmss Z\kern-.4em Z}}
{\lower.9pt\hbox{\cmsss Z\kern-.4em Z}}
{\lower1.2pt\hbox{\cmsss Z\kern-.4em Z}}\else{\cmss Z\kern-.4em Z}\fi}
\def\inbar{\,\vrule height1.5ex width.4pt depth0pt}
\def\IB{\relax{\rm I\kern-.18em B}}
\def\IC{\relax\hbox{$\inbar\kern-.3em{\rm C}$}}
\def\ID{\relax{\rm I\kern-.18em D}}
\def\IE{\relax{\rm I\kern-.18em E}}
\def\IF{\relax{\rm I\kern-.18em F}}
\def\IG{\relax\hbox{$\inbar\kern-.3em{\rm G}$}}
\def\IH{\relax{\rm I\kern-.18em H}}
\def\II{\relax{\rm I\kern-.18em I}}
\def\IK{\relax{\rm I\kern-.18em K}}
\def\IL{\relax{\rm I\kern-.18em L}}
\def\IM{\relax{\rm I\kern-.18em M}}
\def\IN{\relax{\rm I\kern-.18em N}}
\def\IO{\relax\hbox{$\inbar\kern-.3em{\rm O}$}}
\def\IP{\relax{\rm I\kern-.18em P}}
\def\IQ{\relax\hbox{$\inbar\kern-.3em{\rm Q}$}}
\def\IGa{\relax\hbox{${\rm I}\kern-.18em\Gamma$}}
\def\IPi{\relax\hbox{${\rm I}\kern-.18em\Pi$}}
\def\ITh{\relax\hbox{$\inbar\kern-.3em\Theta$}}
\def\IOm{\relax\hbox{$\inbar\kern-3.00pt\Omega$}}

\def\Z{\IZ}

\def\R{\IR}
\def\C{\IC}


\def\d{{\rm d}}
\def\dm{\d^{-1}}
\def\oh{{1\over 2}}\def\un{{\bf 1}}
\def\bz{\bar z}\def\Z{Z^{(N)}}

\def\Ga{\alpha}\def\Gb{\beta}\def\Gc{\gamma}
\def\Gd{\delta}\def\GD{\Delta}\def\Ge{\epsilon}
\def\Gth{\theta}
\def\Gl{\lambda}\def\GL{\Lambda}
\def\ksi{\xi}
\def\Gr{\rho}
\def\Gs{\sigma}\def\GS{\Sigma}

\def\Go{\omega}

\def\mod{{\rm mod\,}}
\def\ch{{\rm ch}}\def\diag{{\rm diag \,}}\def\d{{\rm d}}\def\dm{\d^{-1}}
\def\pd{\partial } \def\pdl{{\partial\over \partial \Gl}}

\def\bra{\langle}\def\ket{\rangle}

\def\P{Pl\"ucker\ }\def\K{Kontsevich}

\lref\Kun{M. \K,
\it Intersection theory on the moduli space of curves\rm,
preprint 1990.}
\lref\Kde{M. \K,
\it Intersection theory on the moduli space of curves and the matrix Airy
function\rm,
Bonn preprint MPI/91-77.}
\lref\Wun{E. Witten,
\it Two dimensional gravity and intersection theory on moduli space\rm,
Surv. in Diff. Geom. {\bf 1} (1991) 243-310.}
\lref\Wde{E. Witten,
\it On the \K\ model and other models of two dimensional gravity\rm,
preprint IASSNS-HEP-91/24}
\lref\Wdep{E. Witten,
\it The $N$ matrix model and gauged WZW models\rm,
preprint IASSNS-HEP-91/26, to appear in Nucl. Phys. B. }
\lref\Wtr{E. Witten,
\it Algebraic geometry associated with matrix models of two dimensional
gravity, \rm
 preprint IASSNS-HEP-91/74.}
\lref\AvM{M. Adler and P. van Moerbeke,
\it The $W_p$--gravity version of the Witten--Kontsevich model\rm,
Brandeis preprint, September 1991
}
\lref\MS{Yu. Makeenko and G. Semenoff,
\it Properties of hermitean matrix model in an external field\rm,
Mod. Phys. Lett. {\bf A6} (1991) 3455-3466.}
\lref\Sa{M. Sato,
\it Soliton equations as dynamical systems on an infinite dimensional
Grassmann manifold\rm,
RIMS Kokyuroku, {\bf 439} (1981) 30-46 \semi
\it The KP hierarchy and infinite-dimensional Grassmann manifolds,\rm
Proc. Symp. Pure Math. {\bf 49 } (1989) 51-66\semi
M. Sato and Y. Sato,
\it Soliton equations as dynamical systems on an infinite dimensional
Grassmann manifold\rm,
in \it Non linear PDE in Applied Science\rm, US-Japan Seminar,
Tokyo 1982, Lecture Notes in Num. Appl. Anal. {\bf 5} (1982) 259-271.}
\lref\IZ{C. Itzykson and J.-B. Zuber,
\it The planar approximation II, \rm
J. Math. Phys. {\bf 21} (1980) 411-421.}
\lref\HC{Harish-Chandra,
\it Differential operators on a semisimple Lie algebra, \rm
Amer. J. Math. {\bf 79} (1957) 87-120.}
\lref\Mu{F.D. Murnaghan,
\it The Theory of Group Representations, \rm
The Johns Hopkins Press, Baltimore, 1938 \semi
H. Weyl, \it The Classical Groups, \rm Princeton Univ. Press,
Princeton, N.J.,  1946.}
\lref\KK{V. Kazakov and I. Kostov, unpublished
\semi I. Kostov,
\it Random surfaces, solvable lattice models and discrete quantum gravity \rm
\it in two dimensions\rm, Nucl. Phys. B (Proc. Suppl.) {\bf 10A} (1989)
295-322. }
\lref\sextet{E. Br\'ezin and V. Kazakov,
\it Exactly solvable field theories of closed strings\rm,
Phys. Lett. {\bf 236} (1990) 144-150\semi
M. Douglas and S. Shenker,
\it Strings in less than one dimension\rm,
Nucl. Phys. {\bf B335} (1990) 635-654\semi
D.J. Gross and A. Migdal,
\it Non perturbative two--dimensional quantum gravity\rm,
Phys. Rev. Lett. {\bf 64} (1990) 127-130.}
\lref\DS{V.G. Drinfeld and V.V. Sokolov,
\it Lie algebras and equations of the Korteweg--de Vries type \rm,
Journ. Sov. Math. {\bf 30} (1985) 1975-2036.}
\lref\Ka{V. Kazakov,
\it The appearance of matter fields from quantum fluctuations of 2D-gravity,
\rm Mod. Phys. Lett. {\bf 4A} (1989) 2125-2139.}
\ifx\answ\bigans\else
\centerline{\bf Contents}\nobreak\medskip{\baselineskip=12pt
 \parskip=0pt\catcode`\@=11
\noindent {0.} {Introduction} \leaderfill{1} \par
\noindent {1.} {Intersection numbers.} \leaderfill{2} \par
\noindent {2.} {The Kontsevich integral.} \leaderfill{6} \par
\noindent \quad{2.1.} {The main theorem} \leaderfill{6} \par
\noindent \quad{2.2.}
{Expansion of $Z$ on characters and Schur functions} \leaderfill{8} \par
\noindent \quad{2.3.}
{Proof of the first part of the Theorem.} \leaderfill{14} \par
\noindent {3.} {From Grassmannians to KdV} \leaderfill{22} \par
\noindent {4.} {Matrix Airy equation and Virasoro highest weight conditions.}
\leaderfill{33} \par
\noindent {5.} {Genus expansion.} \leaderfill{35} \par
\noindent {6.}
{Singular behaviour and Painlev{é} equation.} \leaderfill{42} \par
\noindent {7.}
{Generalization to higher degree potentials} \leaderfill{45} \par
\catcode`\@=12 \bigbreak\bigskip} \fi

\secno=-1
\newsec{Introduction}

\noindent
The study of two--dimensional gravity has uncovered a rich
mathematical structure including Virasoro constraints, KdV flows, $N=2$
twisted supersymmetry, etc. The remarkable contributions of Witten \Wun\
and \K\ \Kde\
to its topological interpretation  have reduced an intersection problem
on the moduli space of curves to the computation of a matrix integral
over $N\times N$ hermitian matrices
\eqn\Oa{
Z(\GL)={\int dM \exp -\tr \({\GL M^2\over 2}-i{M^3\over 6}\)
\over \int dM  \exp -\tr{\GL M^2\over 2}}}
While many aspects of this connection have been already discussed by these
authors\Wde\Wdep, 
our endeavour has been to study this integral in a purely algebraic context,
combining reviews of former work and
new results. We apologize to the expert reader who may skip
sec.~1 where we sketch \K's construction and parts of
sec.~3 which present
a short account of Sato's work on $\tau$--functions. In sec.~2 we study
the integral \Oa\ as an expansion in powers of the traces $\tr \GL^{-r}$.
After taking a suitable large $N$ limit,
we prove the crucial property
that the asymptotic expansion of this integral does not
depend on the even traces $\tr \GL^{-2r}$. This result which followed
in \K's work from topological considerations is derived here in a purely
combinatorial fashion. The arguments although straightforward are unfortunately
rather intricate since they cannot directly apply to the finite $N$ integral,
for which only $N$ of the traces are algebraically independent. The
integral is thus subject to several constraints:

(i) The equations of the KdV hierarchy pertaining to a differential
operator of second order
follow from Sato's work and this independence with
respect to the even traces.
These provide evidence of the equivalence
of \K's model with the one--matrix model considered by Witten in \Wun.

(ii) The Virasoro highest weight constraints follow
from the matrix Airy's equation satisfied by the numerator of \Oa\
(sec.~4) \Wde.

As for the standard matrix models, a systematic genus expansion is
possible, the leading term of which had been obtained by Witten
\Wun\ and from another point of view by Makeenko and Semenoff \MS\
relying on earlier work by Kazakov and Kostov \KK\
(sec.~5). An analysis of
the resulting expressions in a certain singular limit (sec.~6)
provides effective ways to resum families of intersection numbers and
derive explicit fomulae. This step introduces a Painlev\'e equation and
its perturbations.

The integral \Oa\ admits a generalization in which the cubic potential
is replaced by a suitable higher degree polynomial \AvM\Kde.
The corresponding topological interpretation presented in a recent paper
\Wtr\  involves an intersection theory on a  finite covering of moduli
space. On the other hand, it is most likely equivalent to the multimatrix
integrals.
Our previous discussion extends to
these cases without any difficulty of principle (sec.~7) although the
calculations soon become very cumbersome.

\newsec{Intersection numbers.}

\noindent
Witten conjectured in \Wun\ that the logarithm of the
partition function of the general one-matrix model \Ka\sextet,
expressed in terms of
suitable deformation parameters $t_i$ could be expanded as
\eqn\Ia{
\ln Z= \sum_{k_0,\ldots,k_i,\dots}  \langle \tau_0^{k_0}\ldots
\tau_i^{k_i}\ldots
\rangle {t_0^{k_0}\over k_0!}\ldots{t_i^{k_i}\over k_i!}\ldots}
where the bracketed rational coefficients admitted the following
interpretation as intersection numbers. Let $\CM_{g,n}$ be the moduli space
of complex dimension $3g-3+n\ge 0$ of algebraic curves of genus $g$ with $n$
marked points $x_1,\ldots,x_n$ and $\bar{\CM}_{g,n}$ a suitable
compactification. The cotangent spaces at $x_i$ define line bundles
${\CL}_i$ with first Chern class $c_1({\CL}_i)$ interpreted as 2--forms
over $\CM_{g,n}$. 
For integral non-negative $d_f$'s such that
$3g-3+n=\sum_f d_f$ the integral
\eqn\Iaa{\int_{\bar{\CM}_{g,n}} c_1({\CL}_1)^{d_1}\ldots c_1({\CL}_n)^{d_n} }
(independent of the ordering since 2--forms commute and powers are exterior
powers) is a rational positive number when one considers $\CM_{g,n}$ as an
orbifold, \it i.e.\ \rm the quotient of a contractible ball (Teichm\"uller
space) by the mapping class group. If $k_i=\# \{d_f = i,\  i\ge 0\}$ (so
that $\sum_{i\ge 0} i\,k_i =3g-3+n$) then Witten's conjecture was
that $\langle
\tau_0^{k_0}\ldots \tau_i^{k_i}\ldots\rangle$ defined by \Ia\
is given by the
integral \Iaa.
(We have skipped a number of essential technicalities which make
the above definitions sensible).
An intuitive picture of the line bundles $ {\CL}_i$ over $\CM_{g,n}$
is not straightforward
but at least one can see that $\langle\tau_0^3\rangle=1$, since
$\CM_{0,3}$ is
a point! Even to find that $\langle \tau_1\rangle
={1\over 24}$ from the
definition is non trivial. 

These intersection numbers being topological invariants, \K\ has been able
to reduce them to more manageable expressions using a cell decomposition of
$\CM_{g,n}$ inherited from the physicists' ``fat--graph'' expansion of
Hermitian
matrix integrals. One considers connected fat (\it i.e.\ \rm double-line)
graphs with vertices of valency three or more, genus $g$ and $n$ faces
(dual to the $n$ punctures). One assigns to each (double) edge $e$ a positive
length $\ell_e$ and to each face $f$ a perimeter $p_f=\sum_{e\subset f}
\ell_e$, where $e\subset f$ denotes the incidence relations. The set of such
fat graphs with assigned $\ell$'s is a decorated combinatorial model for
$\CM_{g,n}$. Cells have ``dimensions'' over the reals obtained by counting
the
number of independent lengths for fixed values of the perimeters, namely
$E-n$ where $E$ is the number of edges. If $V_p$ denotes the number of
$p$--valent vertices, one has
\eqnn\Ib
$$\eqalignno{E-n-\sum_{p\ge 3}V_p&= 2g-2 \cr
2E= \sum_{p\ge 3}p V_p&\ge  3\sum V_p. &\Ib \cr}$$
Thus
\eqn\Ic{E-n\le 2(3g-3+n)}
with equality (top dimension) if and only if all vertices are trivalent.

Let $k$ be the number of edges bordering a face $f$ and $\ell_1,\ell_2,\ldots,
\ell_k$ be their successive lengths as we circle around the boundary
counterclockwise (the faces come with a positive orientation), up to
cyclic permutation. One introduces the 2--form
\eqn\Id{\Go=\sum_{1\le a<b\le k-1}d\bigg({\ell_a\over p_f}\bigg)\wedge
d \bigg({\ell_b\over p_f}\bigg)}
invariant under a rescaling and a cyclic permutation of the $\ell$'s, which
is the first Chern class of a universal $S_1$ bundle with $k$--gons as
fibers over a combinatorial model for $\CM_{g,n}$. For fixed values of
the perimeters $p_f$ the formula obtained for the intersection numbers is
then
\eqn\Ie{\langle\tau_{d_1}\ldots\tau_{d_n} \rangle=\int \prod_f \Go_f^{d_f} }
the integral being on top-dimensional cells (we omit a discussion of the
compatibility of orientations with the complex structure of $\CM_{g,n}$).
A generating function is obtained by computing
\eqn\If{\int_0^{\infty}\prod_f\Big(dp_f\,e^{-\Gl_fp_f}\Big) \int{(\sum p_f^2
\,\Go_f)^{3g-3+n}\over (3g-3+n)!}}
where the integral sign stands both for the integral over a cell and a
sum over cells of dimension $3g-3+n$ weighted by the inverse of the order
of their automorphism group (orbifold integration).
On the one hand this is
\eqnn\Ig
$$\eqalignno
{\sum_{d_1+\ldots+d_n=3g-3+n}& \langle\tau_{d_1}\ldots\tau_{d_n}\rangle
\int \prod_1^n dp_f\, e^{-\Gl_f p_f} {p_f^{2d_f}\over d_f! } & \Ig\cr
&= 2^{3g-3+n}
\sum_{d_1+\ldots+d_n=3g-3+n} \langle\tau_{d_1}\ldots\tau_{d_n}\rangle
\prod_1^n (2d_f-1)!!\, \Gl_f^{-(2d_f+1)}.
\cr  }$$
On the other hand one can proceed to a direct evaluation using
\eqn\Ih{{1\over (3g-3+n)!} \prod_f dp_f\wedge\bigg(\sum_f
\sum_{{\scriptstyle{1\le a<b\le k_f-1}\atop \scriptstyle{e_a,e_b\subset f}}}
d\ell_a\wedge d\ell_b\bigg)^
{3g-3+n}= 2^{5g-5+2n}\,d\ell_1\wedge\ldots\wedge d\ell_E}
up to an overall
orientation which is henceforth ignored. The computation of the above
Jacobian (which depends on the structure of the graph only through
$g$ and $n$)
is a delicate matter for which we refer to \Kde. Inserting this expression
into the integral one notes that each edge of length $\ell$ is shared by
two faces $f, f'\supset e$, the corresponding integral contributing a
factor ${2\over \Gl_f+\Gl_f'}$, while the ratio of powers of 2 reads
$ 2^{5g-5+2n}/2^{3g-3+n}= 2^{2g-2+n}=2^{E-V}$. By comparison one obtains
\K's main identity with $\Gamma_{g,n}$ the set of all face-labelled
trivalent connected fat--graphs $\Gc$ of genus $g$ and $n$ faces
\eqn\Ii{
\sum_{\Sigma_1^n d_f=3g-3+n} \langle\tau_{d_1}\ldots\tau_{d_n}\rangle
\prod_{f=1}^n{(2d_f-1)!!\over \Gl_f^{2d_f+1}} = \sum_{\Gc\in \Gamma_{g,n}}
{2^{-V}\over \vert {\rm Aut}_{\Gc}\vert} \prod_{e\in \Gc}{2\over
\Gl_f+\Gl_{f'}} }
where as above $V\equiv V_{\Gc}$, $e$ denote the edges of the graph and
$2/ (\Gl_f+\Gl_{f'})$ is the propagator attached to the edge $e$ bordering
$f$ and $f'$.

The right hand side of this  expression is suggestive of the Feynman
expansion of the logarithm of the matrix integral \Oa\
over $N \times N$ Hermitian matrices ($N\to \infty$). Let $\GL$ stand
for a diagonal matrix $\diag(\Gl_0,\ldots,\Gl_{N-1})$ and introduce
the infinite set $t_{\textstyle .}=\{t_0,t_1,\ldots\}$ defined
as\foot{By convention  $(-1)!!=1$.}
\def\tdot{t_{\textstyle .}}
\eqn\Ij{t_i(\GL)=-(2i-1)!! \,\tr \GL^{-2i-1}\ .}
By summing the above expression over $g$ and $n$, one has
\eqnn\Ik
$$\eqalignno{F(\tdot(\GL))&=\sum_{{{\scriptstyle n\ge 1} \atop{\scriptstyle
 d_1,\ldots,d_n\ge 0}} }
{1\over n!} \langle \tau_{d_1}\ldots\tau_{d_n}\rangle \,
t_{d_1}(\GL) \ldots t_{d_n}(\GL) & \Ik \cr
&=\sum_{\Gc\in \Gamma_N}\left({i\over 2}\right)^V
{1\over \vert {\rm Aut}_{\Gc}\vert}\, \prod_{e\in \Gc}{2\over \Gl_f+\Gl_{f'}}
\cr
&\equiv  \ln Z^{(N)}(\GL) \cr}$$
where the last equality is in terms of the Feynman graph expansion of the
integral \Oa\ and $\sum_{\Gamma_N}$ refers to the summation over
fat graphs with $N$ faces
and all possible distinct assignments of variables $\Gl$ to their faces.
We have noted that $(-1)^n=i^V$ since the relation $2E=3V$ for trivalent
graphs implies that the number of vertices $V$ is even $=2p$ and
from $V-E+n=2p-3p+n\equiv 0 \ \mod 2$, it follows that
$n$ is of the parity of $p$. Formula \Oa\
leads to a propagator $2/(\Gl_f+\Gl_{f'})$, while the coupling at
each vertex is ${i\over 2}$. Thus the above reads (after letting
$N\to \infty$)
\eqn\Il{\ln Z(\GL)= \left< e^{\sum_1^{\infty} \tau_d t_d} \right>}
in agreement with \Ia. In the next section, we discuss the precise
mechanism of the $N\to \infty$ limit by which from finite $\GL$
and finitely many independent
$\tr \GL^{-(2i+1)}$  infinitely many independent variables
$t$ are generated.

The first non trivial graphs in the Feynman expansion yield
after some rearrangement
\eqn\In{F=\ln Z= {t_0^3\over 3!}+{t_1\over 24}+{t_0^3 t_1\over 3!}+
\inv{24}\(t_0t_2+{t_1^2\over 2}\)+\ldots}
%
%

\newsec{The Kontsevich integral.}
\subsec{The main theorem}
\noindent
With the normalized measure
\eqn\IIa{d\mu_{\GL}(M) ={dM \exp -\oh \tr \GL M^2 \over
\int dM \exp -\oh \tr \GL M^2 } ,\qquad
dM\equiv \prod_{i=1}^N dM_{ii} \prod_{i<j} d\Re eM_{ij}\,d\Im mM_{ij}}
over $N\times N$ hermitian matrices $M$, $\GL$ standing at first for a
positive definite such matrix, we wish to study the matrix Airy function
\eqna\IIb
$$\eqalignno{\Z=& \int d\mu_{\GL}(M) \exp {i\over 6} \tr M^3 =\sum_{k\ge 0}
\Z_k(\GL) & \IIb a\cr
\Z _k(\GL) =&{(-1)^k\over (2k)!} \int d\mu_{\GL}(M)
\bigg({\tr M^3\over 6}\bigg)^{2k} & \IIb b  \cr}$$
or rather its asymptotic expansion in traces of powers of
$\GL^{-1}$ for large $N$. A generalization is presented in sec.~7.

More precisely the terms $\Z _k (\GL) $ can be expressed as polynomials
with rational coefficients in the variables
\eqn\IIc{\Gth_r= {1\over r} \tr \GL^{-r}.}
Then $\Z_k$ is homogeneous of degree $3k$, if we set
\eqn\IId{\deg \Gth_r = r.}
Only the first $N$ of the $\Gth_r$ are algebraically independent
for an $N\times N$ matrix $\GL$. However \par
\smallskip
\noindent
\def\Gthd{\Gth_{\textstyle .} }
\bf{ Lemma 1 } \it Considered as a function of $\Gth_{\textstyle .}
\equiv\{\Gth_r\}$,
$\Z_k (\GL)$ is independent of $N$ for $3k \le N$ and depends only on
$\Gth_r$, $1\le r\le 3k$. \rm \par
This allows one to define unambiguously the series
\eqn\IIe{Z(\Gthd) = \sum_{k\ge 0} Z_k(\Gthd)}
where
\eqn\IIf{Z_k(\Gthd) = \Z_k (\Gthd)\ , \qquad N \ge 3k}
without any further reference to $N$. The set of variables $\Gth_.$
is denumerable but each $Z_k$ ($Z_0=1$) depends on finitely many of them.
This almost evident lemma follows from eq.~(2.43) 
below.
\bigskip \penalty -300 \noindent
{\bf Theorem 1} (Kontsevich) \it \par
\penalty 600
\eqnn\IIgn{}\par
\noindent (i) 
\qquad\qquad\qquad\qquad {$\displaystyle {\pd Z\over \pd \Gth_{2r}}(\Gthd)
=0 \qquad
r\ge 1$}  
\hfill{\rm \IIgn \it }\par \noindent
(ii) $Z(\Gthd)$ is a $\tau$-function for the Korteweg-de Vries
equation. Namely if
\eqnn\IIh
\eqna\IIhj
$$\eqalignno{t_r &= -(2r+1)!!\,\,\Gth_{2r+1}= -(2r-1)!!\,\,\tr\GL^{-2r-1}\cr
u&=  {\pd^2\over \pd t_0^2 }\ln Z & \IIh \cr
{\textstyle \it then} \qquad\qquad \qquad
{\pd u\over \pd t_1}&= {\pd\over \pd t_0}\bigg( {1\over 12} {\pd^2 u\over
\pd t_0^2 }+\oh u^2 \bigg)& \IIhj a\cr
{\textstyle \it and\ more\ generally} \qquad \qquad
{\pd\over \pd t_n}u &={\pd\over\pd t_0}R_{n+1}.&\IIhj b \cr}$$
\rm
In \IIhj{}, the $R_n$ denote the Gelfand-Dikii differential polynomials
(derivatives are taken with respect to $t_0$)
\eqnn\IIhk
$$\eqalignno{R_2=& {u^2\over 2}+{u''\over 12} \cr
R_3=& {u^3\over 6}+{uu''\over 12}+{u'^2\over 24}+{u^{(4)}\over 240}\cr
R_4=& {u^4\over 24}+{uu'^2\over 24}+{u^2u''\over 24}+{uu^{(4)}\over 240}
+{u'u'''\over 120}+{(u'')^2\over 160}+{u^{(6)}\over 6720}\cr
& \ldots & \IIhk\cr
R_n=& {u^n\over n!}+\cdots\cr}$$
computed from
\eqn\IIl{(2n+1)R'_{n+1}=\inv{4}R'''_n+2u R'_n +u' R_n.}
\medskip
Proof of the first part of theorem 1 follows in \K's approach
from the identities \Ii--\Il,
relying on topological considerations, namely on the introduction of
the combinatorial model for $\CM_{g,n}$. The rest of this section is
devoted to a purely algebraic proof of this result.

%
%
\subsec{Expansion of $Z$ on characters and Schur functions}
\noindent
The integral \IIa\ in the case $N=1$, denoted for short $z(\Gl)$,
\eqn\IIi{z(\Gl)= {\int_{-\infty}^{\infty} dm\, e^{-\oh \Gl m^2+{i\over 6} m^3}
\over \int_{-\infty}^{\infty} dm\,
e^{-\oh \Gl m^2}}=\bigg({\Gl\over 2\pi}\bigg)
^{\oh}  \int_{-\infty}^{\infty} dm\, e^{-\oh \Gl m^2+{i\over 6} m^3} }
admits the asymptotic expansion
\eqna\IIj
$$\eqalignno{
z(\Gl)&= \sum_{k=0}^{\infty}c_k  \Gl^{-3k}&\IIj a\cr
c_k&=\Big(-{2\over 9}\Big)^k {\Gamma(3k+\oh) \over (2k)! \sqrt{\pi}}
&\IIj b \cr} $$
and satisfies the differential equation equivalent to Airy's equation
\eqnn\IIkk
\eqnn\IIk
$$\eqalignno{\big(D^2 & -\Gl^2 \big)z(\Gl)=0 &\IIkk \cr
D&=
- e^{{1\over 3}\Gl^3} \Gl^{\oh}
\big({1\over \Gl} \pdl \big)
e^{-{1\over 3}\Gl^3} \Gl^{-\oh} \cr
&= \Gl +{1\over 2 \Gl^2} -{1\over \Gl}\pdl\ . & \IIk \cr
}$$

For $N$ arbitrary, we can make
a shift of variable $M \to M-i \GL$ in the numerator of
the integral \IIb. This gives
\eqn\IIl{\Z(\GL)= {1\over 2^{{N(N-1)\over 2}} (2\pi)^{{N^2\over 2}}}
\prod_r \Gl_r^{\oh} \prod_{r<s}(\Gl_r+\Gl_s) \exp {\tr {\GL^3\over 3}}
\int dM \exp i\,\tr\bigg( {M^3\over 6}+{M\GL^2\over 2}\bigg)\ ,}
with $\Gl_i$ the eigenvalues of $\GL$.
We then integrate over ``angles'', \it i.e.\ \rm
over the unitary group with the action $M\to UMU^{-1}$.
The result \IZ\ (which we now realize had been established long
before by Harish-Chandra \HC) yields
\eqnn\IIm
$$\eqalignno{
\int dM & \exp i\,\tr  \bigg( {M^3\over 6 }+{M \GL^2\over 2}\bigg)= \cr
&= (2\pi)^{{N^2\over 2 } }\int \prod_{0\le r\le N-1} {dm_r\over (2\pi)^{\oh}}
e^{i\big({1\over 6}m_r^3 +\oh m_r \Gl_r^2\big)}
\prod_{0\le r<s\le N-1} \Big( {m_s-m_r\over i{\Gl_s^2\over 2}-
i{\Gl_r^2\over 2}}\Big) & \IIm \cr
&= (2\pi)^{{N^2\over 2}}{(-1)^{{N(N-1)\over 2}}2^{N(N-1)} \over
\prod_{0\le r<s \le N-1}(\Gl_s^2-\Gl_r^2)} \prod_{0\le r < s \le N-1}
\Big({\pd\over \pd \Gl_s^2}-{\pd\over \pd \Gl_r^2}\Big)
\prod_{r=0}^{N-1} \bigg[e^{-{\Gl_r^3\over 3}}\Gl_r^{-\oh} z(\Gl_r)\bigg]
. \cr}$$
with $m_0, m_1, \cdots, m_{N-1}$ the eigenvalues of the matrix $M$.
We denote the Vandermonde determinant
\eqn\IIn{\left\vert \matrix{x_0^0 & x_0^1& \ldots & x_0^{N-1}\cr
			    x_1^0 & x_1^1& \ldots & x_1^{N-1}\cr
			    \vdots& \vdots & \ddots & \vdots \cr
                            x_{N-1}^0 & x_{N-1}^1& \ldots & x_{N-1}^{N-1}\cr
				}    \right\vert
= \prod_{0\le r < s \le N-1} (x_s-x_r)}
by $\vert x^0, x^1, \ldots , x^{N-1}\vert$, with the understanding that
in each row one substitutes successively $x_0$, $x_1$,$ \cdots $, $x_{N-1}$
for the variable $x$. Inserting \IIm\ into \IIl, we find
%
\eqn\IIp
{\Z(\GL)={
{\vert D^0 z,\ D^1z, \cdots,  D^{N-1}z \vert \over
\vert \Gl^0,\ \Gl^1 ,\cdots, \Gl^{N-1} \vert\  }
}\ ,}
with $D$ defined as in \IIk.
The asymptotic expansion involves only inverse powers of $\GL$. We
set
\eqn\IIq
{x_r =\Gl_r^{-1}\qquad \qquad \Gth_k={1\over k} \sum_{0\le r\le N-1} x_r^k\ ,}
and consider henceforth the function $z$ as given by the formal series
\eqn\IIr
{z(x)=\sum_{0}^{\infty}c_k x^{3k}.}
We have $D^2 z =\Gl^2 z$, and one readily sees that
\eqnn\IIs
$$\eqalignno{D^{2k}z&= \Gl^{2k}z \qquad \mod (D^{2k-1}z, \cdots, D^0 z) \cr
D^{2k+1} z &= \Gl^{2k+1}\bz \quad \mod (D^{2k}z, \cdots, D^0z)\ , &\IIs \cr}$$
with
\eqnn\IIt
$$\eqalignno{
\bz &= {1\over \Gl} D z & \IIt \cr
&= \bigg[1+x^3\bigg(\oh+x{d\over dx}\bigg)\bigg] z
= \sum_0^{\infty} d_k x^{3k}\cr
 d_k&= {1+6k\over 1-6k} c_k. \cr
}$$
As a consequence, we rewrite \IIp\ as
\eqn\IIu{\Z(\GL)={\vert \Gl^0 z,\ \Gl^1\bz,\ \Gl^2 z, \cdots \vert
\over \vert \Gl^0,\ \Gl^1,\ \Gl^2, \cdots, \Gl^{N-1} \vert}\ ,}
where the last term in any row of the upper determinant is $\Gl^{N-1}z$
if $N$ is odd, and $\Gl^{N-1}\bz$ if $N$ is even. Finally we cancel from
numerator and denominator the product $\big( \Gl_0 \cdots \Gl_{N-1}
\big)^{N-1}$, and express $\Z$ in terms of the variables $x_r=\Gl_r^{-1}$ as
\eqn\IIv{\Z = {\vert x^{N-1}z, \, x^{N-2}\bz,\, \cdots \vert \over
\vert x^{N-1},\, x^{N-2},\cdots \vert}\ .}
Then we substitute the expansions of $z$ and $\bz$ to obtain the series
\eqn\IIw{\Z=\sum_{0\le n_0,n_1,\cdots, n_{N-1} }
c_{n_0}^{(0)} c_{n_1 }^{(1)}\ldots c_{n_{N-1}}^{(N-1)}
{\vert x^{3n_0+N-1},\, x^{3n_1+N-2},\cdots, x^{3n_{N-1}} \vert \over
\vert x^{N-1},\, x^{N-2},\cdots, x^{0}\vert}\ ,   }
with the convention that
\eqn\IIx{c_n^{(2p)}=c_n \qquad c_n^{(2p+1)}=d_n. }
%

When the indices $f_0=3n_0$, $f_1=3n_1, \cdots$, $f_{N-1}=3n_{N-1}$
form a non-increasing sequence, we recognize in the above ratio of
determinants the symmetric function in $x_0, \cdots, x_{N-1}$ which
corresponds
to the polynomial character of the general linear group $GL(N)$
specified by the Young tableau with rows of length $f_0$, $f_1, \cdots$,
$f_{N-1}$:
\vskip8truemm
$$\vbox{\offinterlineskip
\def\tv{\vrule height 15pt depth 5pt}
\def\th{\vrule height 0.4pt width 2em}
\def\ca{\hfill}\cleartabs
\+          &&   &\th&\th&\th&\th&\th&\th&\th&\cr
\+ 
&\tv&\ca&\ca&\ca&\ca&\ca&\ca&\ca&\ca&\tv&$\quad l_{N-1}=f_0+N-1$\cr
\+       &  &&   &   &   &   &\th&\th&\th&\cr
\+ 
&\tv&\ca&\ca&\ca&\ca&\ca&\tv&   &   &$\quad l_{N-2}=f_1+N-2$\cr
\+       &  &&   &   &   &   &   &   &\cr
\+        &\tv&\ca&\ca&\ca&\ca&\ca&\tv&   &   &\quad\vdots\cr
\+       &&   &   &   &\th&\th&\cr
\+      &\tv&\ca&\ca&\ca&\tv&   &   &   &   &\quad\vdots\cr
\+       &&   &   &\th&   \cr
\+       &\tv&\ca&\ca&\tv&   &   &   &   &   &\quad\vdots\cr
\+ \cr
\+ 
         & \tv&\ca&\ca&\tv&   &   &   &   &   &\quad $l_0=f_{N-1}$\cr
\+     &&\th&\th&\cr
}\ . $$
%
The latter admits a natural extension outside the standard Weyl
chamber, as an antisymmetric function of the unordered exponents
\eqn\IIy{l_{N-1}=f_0+N-1,\ \cdots \ l_0=f_{N-1}\ .}
Henceforth we refer to this extension when we write this character as
\eqn\IIz{\ch_{l_{N-1},\cdots,l_0}={\vert x^{l_{N-1}},\cdots ,x^{l_0}\vert
\over \vert x^{N-1},\cdots, x^0\vert }\ .}
We wish to express this quantity in terms of the traces $\Gth_k
=\sum_0^{N-1}x_r^k$. This follows from the standard identities for
characters which we now recall \Mu.
Thinking of $X$ as the diagonal matrix $X \equiv \GL^{-1}=\diag
(x_0,x_1,\cdots,x_{N-1})$ we set $s_0=p_0=1$ and for $k\ge 1$
\eqna\IIaa
$$\eqalignno{s_k(X)=&\tr \wedge^k X \qquad
		\ \ \det(\un+u X) =\sum_0^N u^k s_k(X)& \IIaa a\cr
     p_k(X)=&\tr\otimes^k_{{\rm sym}}X \qquad
		{1\over \det(\un-u X)}=\sum_0^{\infty} u^k p_k(X) &\IIaa b\cr
     \Gth_k=&{1\over k}\tr X^k.  &\IIaa c\cr }$$
The quantities $s_k$, $1\le k\le N$, are the elementary symmetric functions
corresponding to the vertical Young tableaux up to $N$ lines, whereas
the $p_k$'s are the traces of symmetric tensor products, \it i.e.\ \rm
they correspond to Young tableaux with only one row.

We have from the definition
\eqn\IIab{\exp \sum_1^{\infty}u^n \Gth_n(X)=\sum_0^{\infty}u^k p_k(X)\ ,}
which expresses the $p_k$'s as homogeneous polynomials of degree $k$ of
the $\Gth_n$'s ($\deg \Gth_n=n$), ignoring the relations among traces. This
justifies the definition of (elementary) Schur functions $p_n$ via
\eqn\IIac{\exp \sum_1^{\infty}u^n \Gth_n =\sum_0^{\infty}u^k p_k(\Gthd)\ ,}
without reference to any $N\times N$ matrix, and where now the $p$'s are
functions of the $\Gth$'s. When both $p$ and $\Gth$ refer to the same matrix
$X$ we recover the previous definitions \IIaa{} and \IIab. Explicitly
we write
\eqn\IIaca{p_r(\Gthd)=\sum_{\nu_1+2\nu_2+\ldots=r}
{\Gth_1^{\nu_1}\over \nu_1!} {\Gth_2^{\nu_2}\over \nu_2!}\ldots\ .}
Eq.\IIac\ entails
\eqn\IIad{{\pd p_r(\Gth_.)\over \pd\Gth_k}= {\pd^k p_r(\Gth_.)
\over \pd\Gth_1^k} =p_{r-k}(\Gth_.)\ , }
where $p_n$ vanishes for $n<0$.
When expanding the matrix elements along successive columns,
Cauchy's determinental formula
\eqn\IIae{\det\left\vert {1\over 1-x_r y_s}\right\vert_{0\le r,s\le N-1}
= {\vert x^{N-1},\cdots,x^0\vert \, \vert y^{N-1},\cdots,y^0\vert
\over \prod_{r,s}(1-x_r y_s) } }
yields
\eqn\IIaf{\sum_{l_0,\cdots,l_{N-1}}  y_0^{l_{N-1}} \cdots y_{N-1}^{l_0}
{\vert x^{l_{N-1}},\cdots,x^{l_0}\vert\over \vert x^{N-1},\cdots,x^0\vert}
= {\vert y^{N-1},\cdots,y^0\vert\over \prod_{r,s} (1-x_r y_s)}\ .}
Therefore if $X\equiv \diag(\{x_r\})$, then
\eqn\IIag{\ch_{l_{N-1},\cdots,l_0}(X)= {\rm coeff.\ of\ } y_0^{l_{N-1}}
\cdots y_{N-1}^{l_0} \   {\rm in \ }
\left\vert   \matrix{{y_{0}^{N-1}\over \det(\un-y_{0}X)} & \ldots &
{y_{0}^{0}\over \det(\un-y_{0}X)} \cr
\vdots &\ddots   & \vdots \cr
{y_{N-1}^{N-1}\over \det(\un-y_{N-1}X)} & \ldots
& {y_{N-1}^{0}\over \det(\un-y_{N-1}X)}  \cr }\right\vert\ . }
Expanding $\Big[\det(\un-yX) \Big]^{-1}$ according to \IIaa{b}, we obtain
the classical formula (Jacobi-Schur)
\eqn\IIah{\ch_{l_{N-1},\cdots,l_0}(X) =
{\vert x^{l_{N-1}},\cdots,x^{l_0}\vert\over \vert x^{N-1},\cdots,x^0\vert}
=\left\vert\matrix{
p_{l_{N-1}-(N-1)}(X) &\ldots p_{l_{N-1}-1}(X) &p_{l_{N-1}}(X) \cr
p_{l_{N-2}-(N-1)}(X) &\ldots p_{l_{N-2}-1}(X) &p_{l_{N-2}}(X) \cr
\vdots & &\vdots \cr
p_{l_0-(N-1)}(X) &\ldots p_{l_0-1}(X) &p_{l_0}(X) \cr } \right\vert\ ,}
valid for any ordered or unordered sequence $l_{N-1}, \cdots, l_0$. Terms
along the diagonal read $p_{f_0},p_{f_1},\cdots,p_{f_{N-1}}$, and indices
increase (decrease) by successive units as one moves from a diagonal
term to the right (left). We abbreviate this expression as
\def\bullet{\star}
\eqn\IIai
{\ch_{N-1+f_0,\cdots,f_{N-1}}(\Gthd)= \left\vert\matrix{
p_{f_0} & \bullet&\ldots&\bullet \cr
\bullet& p_{f_1}& \ldots& \bullet \cr
\vdots&\vdots &\ddots &\vdots \cr
\bullet&\bullet &\ldots &p_{f_{N-1}} \cr }
\right\vert\ ,}
substituting for the elementary Schur polynomials their expressions 
in terms of the variables $\Gth_.$. We conclude that
\eqn\IIaj{
\Z=\sum_{n_0,\cdots,n_{N-1}} c_{n_0}^{(0)} c_{n_1}^{(1)} \cdots
c_{n_{N-1}}^{(N-1)} \left\vert\matrix{
p_{3n_0}& & \cr
&\ddots & \cr
&  &  p_{3n_{N-1}} \cr }\right\vert}
yields an expression of $\Z$ in terms of the infinitely many variables
$\Gthd$ (which can henceforth be treated as independent). It follows from
eq.~\IIaj\ that
\eqn\IIak{\Z_k=\sum_{n_0+\cdots+n_{N-1}=k} c_{n_0}^{(0)} c_{n_1}^{(1)} \cdots
c_{n_{N-1}}^{(N-1)} \left\vert\matrix{
p_{3n_0}& & \cr
&\ddots & \cr
&  &  p_{3n_{N-1}} \cr }\right\vert  }
where each character is of degree $3k$. This is obvious for the
diagonal term and, as one readily ascertains, holds also for non-diagonal
terms.

We are now in position to prove the lemma, which is trivially true for
$\Z_0=1$. Suppose $0<3k\le N$ and for a given term in \IIak\ let $\Gd$
be its ``depth'', \it i.e. \rm the
smallest integer $\le N$ such that $r\ge \Gd \Rightarrow
n_r=0$. From \IIai\ it follows that the corresponding
term reads
%
$${ c_{n_0}^{(0)} c_{n_1}^{(1)} \cdots
c_{n_{\Gd-1}}^{(\Gd-1)} \left\vert\matrix{
p_{3n_0}& & \cr
&\ddots & \cr
&  &  p_{3n_{\Gd-1}} \cr }\right\vert \qquad; \qquad n_0+\cdots+
n_{\Gd-1}=k .}$$
The last column of the $\Gd\times \Gd$ determinant reads
$\big(p_{3n_0+\Gd-1},\cdots,p_{3n_{\Gd-1}} \big)^T$ where the subscripts
are $\Gd$ positive integers with a sum equal to $3k+\sum_0^{\Gd-1}r$.
If $3k<\Gd$, this is smaller than the sum of the first $\Gd$ positive
integers. From Dirichlet's box principle, two of the subscripts among
$3n_0+\Gd-1$, $3n_1+\Gd-2,\cdots$, $3n_{\Gd-1}$ have to be equal, which
results in two identical lines in the determinant which therefore vanishes.
Hence $\Gd$ has to be smaller than or equal to $3k$, showing that
\eqn\IIam
{N\ge 3k \quad\Longrightarrow\quad \Z_k=Z^{(3k)}_k\equiv Z_k}
which concludes the proof and allows a definition of the formal series
$Z$, without reference to $N$.
%
%
\subsec{Proof of the first part of the Theorem.}
%
(i) The first part of the theorem will be proved for each $Z_k$ which we
take equal to $Z^{(3k)}_k$. Differentiating each line successively in the
determinental characters using the crucial formula \IIad\ we find
\eqnn\IIan
$$\eqalignno{
2r>3k & \qquad {\pd Z_k\over \pd \Gth_{2r} }= 0& \IIan \cr
2r\le 3k & \cases{ {\pd Z_k\over \pd \Gth_{2r}}&=\
$ \sum_{s=0}^{3k-1}Z_{k,(s)} $ \cr
Z_{k,(s)}&= $\sum_{n_0+\cdots+n_{3k-1}=k} c_{n_0}^{(0)} c_{n_1}^{(1)} \cdots
c_{n_{3k-1}}^{(3k-1)} \left\vert\matrix{
p_{3n_0} & &\cr
    \quad  \ddots& & \cr
&  p_{3n_s-2r}&  \cr
& \quad \ddots & \cr
& &  p_{3n_{3k-1}} \cr }\right\vert $ \cr
}
\cr}$$
where subscripts in the $s$-th row of each determinant have been decreased
by $2r$ units. For $0\le s\le 3k-1-2r$ the subscripts in line $s$ and
$s+2r$ only differ by the interchange of $n_s$ and $n_{s+2r}$. The
determinental character is therefore antisymmetric in the interchange of
indices $n_s$ and $n_{s+2r}$ whereas in the product of $c$'s, due to
\IIx\ $\ldots c_{n_s}^{(s)}\ldots c_{n_{s+2r}}^{(s+2r)}\ldots$ $\equiv
\ldots c_{n_s}^{(s)}\ldots c_{n_{s+2r}}^{(s)}\ldots$ is symmetric in these
indices. As a consequence, $Z_{k,(s)}$ vanishes for $s\le 3k-1-2r$ and we
need only consider terms with $s>3k-1-2r$, \it i.e.\ \rm when the derivative
acts on
one of the last $2r$ lines and we cannot use the above argument relying
on the periodicity $c_n^{(r)}=c_n^{(r+2)}$.

(ii) Therefore fix $s$ such that $3k-2r\le s\le 3k-1$.
The only possibly non-vanishing terms in $Z_{k,(s)}$ are those whose
depth $\Gd$ defined as above to be the
smallest integer such that $r\ge \Gd \Rightarrow n_r =0$, satisfies the
inequality $3k-2r\le s \le \Gd-1 \le 3k-1$. In this case they read
\eqn\IIao
{
 c_{n_0}^{(0)} c_{n_1}^{(1)} \cdots
c_{n_{\Gd-1}}^{(\Gd-1)} \left\vert\matrix{
p_{3n_0}& & & \cr
      &\ddots  & & \cr
& & p_{3n_s-2r} & \cr
& & \quad \ddots & \cr
    & &    &  p_{3n_{\Gd-1}} \cr }\right\vert \quad ,
n_{\Gd-1}>0,n_0+\ldots+n_{\Gd-1}=k \ .}
The $\Gd$ indices of the $p$'s in the last column
of the determinant before subtracting $2r$ from the indices of
the $s$-th row are all positive integers and have  a sum equal to $3k-\Gd+
\sum_1^{\Gd}t$, where $0\le 3k-\Gd\le 2r-1$. We now make use of the following
\par
\smallskip
\penalty -400
\noindent \bf{Lemma 2.} 
\par
\penalty 600
\noindent \it If from a set of $\Gd$ positive integers, with sum exceeding
the one of the first $\Gd$ positive integers by an amount $\GD\ge 0$, one
decreases one by $\GD'>\GD$, then in the new sequence two terms
coincide or one is a non positive integer. \rm

Think of the original set as occupied integral levels. Let $r_0+1$ be the
first unoccupied one ($r_0\ge 0$) and $r_1$ the greatest occupied one. It
follows from the hypothesis that $r_1-r_0 \le \GD+1$. If one decreases
one element of the set by an amount $\GD'\ge\GD+1$, it therefore becomes
less than or equal to $r_0$, which proves the lemma.

Applying this result to the above circumstance ($\GD=2r-1, \GD'=2r$),
we deduce that the only possibly non-vanishing terms in $Z_{k,(s)}$,
$3k-2r\le s\le 3k-1$ occur when $3n_s=2r-(\Gd-1-s)$ with
$0\le \Gd-1-s\le 2r-1$. Thus $2r-(\Gd-1-s)$ takes the possible
values $1,\ldots, 2r$. If $r=1$ this is never a multiple of 3. Thus we have
already obtained
\eqn\IIap{{\pd Z\over \pd\Gth_2}=0.}
We can even say more. Let $a$ be the integral part of $(\Gd-1)/3$
and consider in the last column starting from the bottom the $a+1$
positive subscripts
$$ 3n_{\Gd-1},3(n_{\Gd-1-3}+1),\ldots, 3(n_{\Gd-1-3a}+a)\ . $$
For the corresponding character to be non-zero, these have to be all
distinct. Hence their sum is larger than or equal to
$3\sum_0^a (\Ga+1)$. The inequality
\eqn\IIaq{3\sum_{\Ga=0}^a \big( n_{\Gd-1-3\Ga}+\Ga\big)\ge 3\sum _0^{a+1}
\Ga }
implies that $\sum_{\Ga =0}^a n_{\Gd-1-3\Ga} \ge a+1$. Should a
non-vanishing term arise in $Z_{k,(s)}$, there would exist an index
$n_s$ such that from the preceding observation
\eqn\IIar{3n_s+\Gd-1-s = 2r}
%
with
\eqn\IIas{3k-2r \le s \le \Gd-1\ .}
Thus  $3n_s+\Gd-1=2r+s\ge 3k$, or equivalently $n_s+a\ge k$. If
$2r$ is not a multiple of 3,  \it i.e.\ \rm
\eqn\IIat{r\ne 0 \quad \mod 3\ ,}
it follows that
\eqn\IIatt{\Gd-1-s\ne 0 \ \mod 3\ .}
This means that the subscript $3n_s+\Gd-1-s$ does not belong to
the sequence $3n_{\Gd-1-3\Ga} +3\Ga$. Since the sum of all $n$'s is
$k$ we should have %
\eqn\IIau{k\ge \sum_0^a n_{\Gd-1-3\Ga}+n_s\ge a+1+n_s}
whereas from the above $a+1+n_s\ge k+1$, a contradiction. Thus in
general
\eqn\IIav{r\ne 0\ \ \mod 3 \qquad {\pd Z\over \pd \Gth_{2r}}=0.}

(iii) The remaining cases are those for which we take a derivative with
respect to $\Gth_{6r} $. We shall need a relation between the
coefficients $c_n$, $d_n$ which did not play any specific role
until now.
The series  $z(x)$ of \IIr\
%
%
is a solution of the following differential equation in the variable $x$
\eqn\IIax{x^4 z''+2(2x^3+1) z'+{5\over 4}x^2 z=0 \ ,}
while $\bz(x)$ of \IIt\ satisfies
\eqn\IIay{\bz(x)=\sum_0^{\infty}d_n x^{3n}=\bigg(1+{x^3\over 2}\bigg)z
+x^4 z'.}
\smallskip
\noindent {\bf Lemma 3}
\eqn\IIaz{z(x)\bz(-x)+z(-x)\bz(x)=2}
\medskip
To see this, substitute $\bz$ in terms of $z$ and compute the derivative
with respect to $x$, making use of the differential equation for
both $z(x)$ and $z(-x)$, to find that it vanishes.

In terms of the series expansions this reads
\eqn\IIbb{\sum_{s=0}^{2n}(-1)^s c_{2n-s}\, d_s=0 \qquad \qquad n>0.}

It was convenient to use generalized characters until now,  but we can
also recast  the expansion of $Z$  in terms of standard characters
indexed by Young tableaux $Y$ ($f_0\ge f_1\ge \ldots \ge f_{\Gd-1}>0$)
at the price of having more complicated coefficients.
{}From eqs.~\IIah\ and \IIaj\ this reads
\eqn\IIbc{Z=\sum_{Y,\ \vert Y \vert =0, \mod 3} \xi_Y\, \ch_Y(\Gthd)}
with
\eqn\IIbd{\ch_Y(\Gthd)= \det \{p_{f_i+j-i}(\Gthd)\}\qquad
0\le i,j\le \Gd-1}
and
\eqnn\IIbe
$$\eqalignno{
\xi_Y&= \det \xi_{i,j}& \IIbe \cr
\xi_{i,j}&= \cases{0 \qquad & if \ $ f_i+j-i \ne 0 \ \mod 3$ \cr
c_{{1\over 3}(f_i+j-i)}^{(j)} \qquad & if \ $ f_i+j-i =0 \ \mod 3 $\cr
}\ . \cr
}$$
In the above, $i$ ($j$) is the row (column) index. The diagonal
subscripts in $\ch_Y$ are no longer multiples
of 3, instead running down the diagonal
they form a non-decreasing sequence, the last one being positive (and
$\Gd$ being the number of rows of the corresponding Young tableau).
The quantity $Z_k$ is obtained by restricting the sum to $\vert Y\vert =3k$.

When taking a derivative with respect to $\Gth_{6r}$, the first
contributing term is $Z_{2r}$. (Recall that terms in $Z_k$ are homogeneous
of degree $3k$). Let us therefore investigate first
${\pd Z_{2r}\over\pd \Gth_{6r}}$.
The only Young tableaux such that $\vert Y \vert=6r$
for which
$ \pd \ch_Y/ \pd \Gth_{6r}\ne 0$
are of the ``Fermi-Bose'' type $(t+1) (1)^s$, $s+t+1=6r$
\vskip8truemm
\def\tv{\vrule height 15pt
 depth 5pt}
\def\th{\vrule height 0.4pt width 2em}
\def\ca{\quad}\cleartabs
\setbox11=\hbox{$\vbox{\offinterlineskip
\+          &&   &\th&\th&\th&\th&\th&\th&\th&\cr
\+ 
&\tv&\ca&\tv\ca&\tv\ca&\tv\ca&\tv\ca&\tv\ca&\tv\ca&\tv\ca&\tv&\cr
\+       &  &&\th&\th&\th&\th&\th&\th&\th&\cr
}$}
\setbox12=\hbox{$\vcenter{\offinterlineskip
\+       &\tv&\ca&\ca&\tv&   &   &   &   &   &\cr
\hrule
\+       &\tv&\ca&\ca&\tv&   &   &   &   &   &\cr
\hrule
\+ 
         & \tv&\ca&\ca&\tv&   &   &   &   &   & \cr
\hrule
}$}
\setbox22=\hbox{$\left.\vbox to \ht12{}\right\}$}
\setbox33=\hbox{$s$}
\setbox10=\hbox to 14em {\downbracefill}
\eqnn\equerr
$$\vbox{\offinterlineskip\halign{
#& \qquad #& \hfill # \cr
\hbox to 14em{\hfill\quad$ t+1$ \quad\hfill}\cr
\noalign{\vskip 3mm}
\box10 \cr
\noalign{\vskip 3mm}
\box11 \cr
\box12\box22\box33 \cr
}}\eqno\equerr$$
\vskip8truemm
\noindent in which case
$ \pd \ch_Y/ \pd \Gth_{6r}= (-1)^s$. Applying formula
\IIbc\ we get
\eqnn\IIbf
$$\eqalignno{{\pd Z_{2r}\over \pd \Gth_{6r}}=& 2\sum_{s=0}^{2r-1}(-1)^s
\left\vert \matrix{c_{2r-s}& d_{2r-s+1} & \ldots & c_{2r}^{(s)}\cr
                    1      & d_1        & \ldots & c_s^{(s)} \cr
		    \vdots & \ddots           & \ddots & \cr
		    0      & \ldots     &    1   & c_1^{(s)}\cr  }
\right\vert\cr
&+ \sum_{s=0}^{2r-1}(-1)^s
\left\vert \matrix{d_{2r-s}& c_{2r-s+1} & \ldots & c_{2r}^{(s+1)}\cr
                    1      & c_1        & \ldots & c_s^{(s+1)} \cr
		    \vdots &     \ddots       & \ddots & \cr
		    0      & \ldots     &    1   & c_1^{(s+1)}\cr  }
\right\vert\ .& \IIbf \cr
}$$
Expanding each of these determinants along the first column
(where $c$ and $d$ with negative index are set equal to zero), we find
\eqnn\IIbg
$$\eqalignno{
{\pd Z_{2r}\over \pd \Gth_{6r}}=& \sum_{j=0}^r\[ (1-3j)
c_{2j-1}\GD_{2r-(2j-1)} +3j c_{2j}\GD_{2r-2j}  \] \cr
+&\sum_{j=1}^r \[3j d_{2j} \bar\GD_{2r-2j }-(3j-2)
d_{2j+1}\bar\GD_{2r-2j-1}\]\ ,
& \IIbg \cr
}$$
where $\GD_0=\bar\GD_0=1$ and
\eqn\IIbh
{\GD_s=\left\vert\matrix{d_1&c_2&\ldots&c_s^{(s)}\cr
                           1&c_1&\ldots&c_{s-1}^{(s)}\cr
			   \vdots & \ddots& \ddots &\cr
			   0& \ldots & 1 & c_1^{(s)}\cr }\right\vert
\qquad \bar\GD_s=
\left\vert\matrix{c_1&d_2&\ldots&c_s^{(s+1)}\cr
                           1&d_1&\ldots&c_{s-1}^{(s+1)}\cr
			   \vdots & \ddots& \ddots &\cr
			   0& \ldots & 1 & c_1^{(s+1)}\cr }\right\vert. }
Taking into account the identity \IIbb\ satisfied by the coefficients $c$
and $d$ we readily see by a recursive argument that
$\GD$ and $\bar\GD$ reduce to
\eqn\IIbi{\GD_s= d_s \qquad\qquad \bar\GD_s=c_s}
Thus
\eqn\IIbj{{\pd Z_{2r}\over \pd\Gth_{6r}}=(3r-1) \sum_{j=0}^{2r} (-1)^j
c_j d_{2r-j} =0.}

(iv) It remains finally to examine
$$ {\pd Z_{2r+k}\over \pd\Gth_{6r}}\qquad\qquad k>0$$
Let us look at the expression \IIan\ with the required modification
$k\to k+2r$, $2r\to 6r$. As before when taking derivatives of the row
of index $s$ we need only take into account those terms for which $s\ge 3k$,
the others vanishing due to the antisymmetry of the characters. This being
assumed we consider for fixed $s$ a specific term in the sum \IIan\ with
character of depth $\Gd>3k$ so that we can omit the rows and columns
of label larger than $\Gd-1$ in the computation of the corresponding
determinental character. The labels in the last column before derivation
are $3n_0+\Gd-1,\ldots$,$3n_{\Gd-1}$, no pair of them equal. According to
a previous analysis using lemma 2, to get a non-trivially vanishing
derivative, the quantity $3n_s+\Gd-1-s$ ($s\ge 3k$) has to
equal $6r$ . This means that
$\Gd-1-s=3\Gs$, and $n_{\Gd-1-3\Gs}=2r-\Gs$. Since by definition
$n_{\Gd-1}>0$, from the preceding reasoning we must have
$\sum_{\Ga=0}^{\Gs-1}n_{\Gd-1-3\Ga}
\ge \Gs$, if  this sum is non-empty  (\it i.e.\ \rm$\Gs>0$). Then
\eqn\IIbk{\sum_{\Ga=0}^{\Gs} n_{\Gd-1-3\Ga}\ge 2r\ ,}
an inequality which remains obviously true when $\Gs=0$, in which case
$n_{\Gd-1}=2r$. \it A fortiori \rm
\eqn\IIbl{\sum_{\Gr=3k}^{\Gd-1}n_{\Gr}\ge 2r\ ,}
since the latter sum includes the previous one ($\Gd-1-3\Gs\ge 3k$). Since
$\sum_0^{\Gd-1}n_{\Gr}=2r+k$ from the homogeneity property of
$Z_{2r+k}$, we have the complementary inequality
\eqn\IIbm{\sum_0^{3k-1}n_{\Gr}\le k.}
Both inequalities must in fact be equalities. Indeed among the integers
of the form $\Gd-1-3\Ga$ ($\Ga\ge 0$) such that $\Gd-1-3\Ga\ge k$,
consider the largest one, say $\Gb$, for which $n_{\Gd-1-3\Ga}> 0$. We
have $\Gb\ge \Gs$ and again appealing to a previous reasoning
%
\eqn\IIbn{\sum_{{{\scriptstyle 0\le\Gr<\Gd-1-3\Gb}\atop{\scriptstyle
\Gr=\Gd-1\,\mod 3}}}n_{\Gr}\ge k\ , }
since there are at least $k$ integers equal to $\Gd-1\,\mod 3$ between 0 and
$3k-1<\Gd-1-3\Gb$. Hence \par\noindent
(i) $\sum_{\Ga=0}^{\Gs}n_{\Gd-1-3\Ga}$ has to equal $2r$ otherwise
we would violate homogeneity;\par\noindent
(ii) $\Gb$ has to be equal to $\Gs$ for the same reason (no other
$n_{\Gd-1-3\Ga}$ except those entering the previous sum are $>0$);\par
\noindent(iii) and finally should $n_{\Gr_0}$, $\Gr_0\ge 3k,\ \Gr_0\ne \Gd-1
\ \mod 3$, be positive, then again $\sum_{{0\le\Gr\le\Gr_0\atop\Gr=\Gr_0\,
\mod 3}}n_{\Gr}$
would be larger than $k$ (since the sum includes $\Gr_0\ge 3k$), a
contradiction. We conclude that we can replace \IIbm\ and \IIbn\ by equalities.
This means that we can write
\eqn\IIbo
{{\pd Z_{2r+k}\over\pd\Gth_{6r}}=\sum_{s\ge 3k}
\sum_{{{\scriptstyle n_0+\cdots+n_{3k-1}=k}\atop{
\scriptstyle n_{3k}+\ldots+n_{6r+3k}=2r} }}
c_{n_0}^{(0)} \cdots c_{n_{6r+3k-1}}^{(6r+3k-1)}
\left\vert\matrix{p_{3n_0} & & & \cr
                         &\ddots  & &\cr
			 & & p_{3n_s-6r} & \cr
			 &  &\quad \ddots & \cr
			 &  & & p_{3n_{3k+6r-1}}  }\right\vert }
Let us group together all contributions corresponding to a fixed choice
of indices $n_0,\ldots,n_{3k-1}$ with sum equal to $k$. Taking into
account the antisymmetry in the last $6r$ rows we see that the analysis
reduces to our previous computation of $\pd Z_{2r}/ \pd\Gth_{6r}$. To be
precise, the column labelled $3k$ will correspond to coefficients labelled
$c$ or $d$ according to $k$ even or odd so that  if $k$ is even
\eqn\IIbp
{{\pd Z_{2r+k}\over\pd\Gth_{6r}}=
\sum_{n_0+\cdots+n_{3k-1}=k}
c_{n_0}^{(0)} \cdots c_{n_{3k-1}}^{(3k-1)}
\left\vert\matrix{p_{3n_0} & & \cr
                         &\ddots &\cr
			 & &  p_{3n_{3k-1}}  }\right\vert
\,{\pd Z_{2r}\over\pd\Gth_{6r}}\ . }
If $k$ is odd, we get the same formula with
$\pd Z_{2r}/\pd\Gth_{6r} $ replaced by the same expression with
$c$'s and $d$'s interchanged. In both cases the expression vanishes
since $c$'s and $d$'s play a symmetric role in the vanishing of
$\pd Z_{2r}/\pd\Gth_{6r}$ as it relies on the identity \IIaz\ invariant
in the interchange of $z$ and $\bz$.

We have at last fully proved the first part of Kontsevich's
theorem from a purely algebraic standpoint
\eqn\IIbq{{\pd Z_{k}\over\pd\Gth_{2r}}=0}
Even though the proof appears a little long, the steps are completely
elementary relying on the second order differential equation satisfied
by $z$ and Weyl antisymmetry of characters. Without repeating in detail
each step, it will go through in the generalized case considered in sec.~7.

It may also be worth remarking that by retracing the above discussion,
this property is very likely to imply the (Airy-like)
differential equation. We now turn to the second part of the theorem.

%
\newsec{From Grassmannians to KdV}

\noindent Expert readers will have recognized the connection between the
expansion \IIbc\ of $Z$ and Sato's approach to soliton equations and $\tau$-
functions. The latter relies on a clever reinterpretation of the familiar \P
relations of projective geometry in terms of properties of associated
(pseudo-)differential operators \Sa. In more physical terms, this involves
the characterization of those submanifolds
that correspond to pure Slater determinants (as opposed to their
linear combinations) in a many body fermionic space.

\def\ksi{\underline{\xi}}\def\ue{\underline{e}}
Let us begin with a short review of the subject following Sato. Let $V$ be
a vector space of dimension $N$ equipped with a basis $\ue_0,\cdots,
\ue_{N-1}$.
The field of constants is arbitrary but one may think of $\R$ or $\C$. A
linear subspace generated by $m$ vectors $\ksi^{(0)},\cdots,\ksi^{(m-1)}$
is intrinsically described by the antisymmetric multivector
\eqn\IIIa{\ksi^{(0)}\wedge\cdots\wedge\ksi^{(m-1)}=\sum_{0\le
l_0\le\cdots\le l_{m-1}\le
N-1} \xi_{l_0,\cdots,l_{m-1}} \ue_{l_0}\wedge\cdots \wedge \ue_{l_{m-1}}}
with antisymmetric components
\eqn\IIIb{\xi_{l_0,\cdots,l_{m-1}} =\det \xi_{l_i,j}\ ,\qquad
0\le i,j\le m-1\ , \qquad \ksi^{(k)}=\xi_{i,k}\ue_i}
All relations to be written being homogeneous we may consider the above
quantities as being homogeneous components of the corresponding $m-1$
dimensional linear subspace in the projective space $PV(N-1)$
of dimension $N-1$. A familiar case is the description of lines ($m=2$)
<in projective three space ($N=4$). An $(m-1)$-dimensional subspace in
$PV(N-1)$ depends on $m(N-m)$ parameters\foot{This is $m\times N$, the
number of components of the vectors $\ksi^{(k)}$, minus $m^2$, the dimension
of the linear group $GL(m)$ acting linearly on this vector without
modifying the subspace.} (four in the above example, for instance the
intersections of the line with two planes) while the number of 
coordinates in \IIIa\ (taking into account homogeneity) is ${N\choose m}-1$
(\it i.e. \rm 5 in the example). They must therefore satisfy some
(non-linear but homogeneous) relations.  These are the \P relations. In the
aforementioned example this is the classical quadratic relation
expressing that the geometry of lines in the projective 3--dimensional space
is equivalent to the geometry of points on a quadric in 5--dimensional
projective space.

\def\uet{\underline{\eta}}
To derive typical \P relations, we demand that a linear combination
\eqn\IIIc{\uet=\sum x_k\, \ksi^{(k)}}
lies in the subspace generated by the $\ksi$'s, \it i.e.\ \rm that
\eqnn\IIId
$$\eqalignno{0=\uet\wedge \ksi^{(0)}\wedge \cdots \ksi^{(m-1)} &
=\sum_{0\le l_0 \cdots \l_m\le N-1} \tilde\xi_{l_0,\cdots,l_m}\, \ue_{l_0}
\wedge \cdots \wedge \ue_{l_m} & \IIId \cr
\tilde\xi_{l_0,\cdots,l_m} &= \sum_{i=0}^m (-1)^i \sum_{r=0}^{m-1}
x_r\, \xi_{l_i}^{(r)} \xi_{l_0,\cdots,\widehat{l_i},\cdots,l_m}\ .\cr
}$$
Choose the coefficients $x_r$ as the minors in the last line of
\eqn\IIIe{\xi_{k_0,\cdots,k_{m-2},l}\equiv
\left\vert\matrix{ \xi_{k_0}^{(0)} & \ldots & \xi_{k_0}^{(m-1)} \cr
                   \vdots & & \vdots \cr
		   \xi_{k_{m-2}}^{(0)}& \ldots & \xi_{k_{m-2}}^{(m-1)}\cr
		   \xi_{l}^{(0)}& \ldots & \xi_{l}^{(m-1)}\cr
                }  \right\vert
=\sum_{r=0}^{m-1} x_r\, \xi_l^{(r)}}
which only depend on the choice of $k_0,\cdots,k_{m-2}$ and not on the
index $l$. Hence we get the \P relations in the form
\eqn\IIIf{\sum_{i=0}^m \ksi_{k_0,\cdots,k_{m-2},l_i}\,\ksi_{l_0,\cdots,
\widehat{l_i}, \cdots,l_m}=0 .}
The reader might have fun to find the relations among these relations
and so on. In any case by turning the argument around these relations
do characterize $m$--dimensional vector subspaces in $V$ or the
$(m-1)$--dimensional ones in $PV$ which form the $(N,m)$ Grassmannian
(obviously not a vector space but rather an intersection of quadrics).

For our purposes we will need a generalization of \IIIe\ which follows from
the observation that we could equally well replace $\eta$ by some combination
\eqn\IIIg{\uet'=\sum_{k<k'}x_{k,k'}\, \ksi^{(k)}\wedge \ksi^{(k')}}
(or for that matter by any higher superposition of exterior products).
Leaving the general case aside, we also have
\eqn\IIIh{\sum_{i<j} (-1)^{i+j-1} \xi\dup_{k_0,\cdots,k_{m-3},l_i,l_j}
\, \xi_{l_0,\ldots,\widehat{l_i},\cdots,\widehat{l_j},\ldots,l_{m+1}}}
and so on.

The relevance of this discussion to the present problem arises from the
determinental expressions for the function $Z$ as expressed in eqs.~\IIw\
and \IIbc. Indeed it was Sato's idea to associate to points of
the Grassmannian a $\tau$--function obtained by replacing exterior
products of basis vectors by the corresponding antisymmetric generalized
Schur functions. In our case one can view
vector subspaces as those generated by the
functions $z,Dz,D^2z,\cdots$ so that \K's integral appears as a
realization of Sato's idea.
The task is now to translate equivalents of the \P
relations in terms of $Z$.

It will be easier to state this in the finite $N$ case we started from,
since by letting $N$ become arbitrarily large we will recover the required
results term by term in the asymptotic series. Thus we return to formula
\IIaj\ understanding by $p_n(\Gthd)$ the unconstrained Schur functions
which we recast in the following form
\eqn\IIIi{\Z= \left\vert \matrix{
{\pd^{N-1} f_{N-1}\over \pd \Gth_1^{N-1}}&\ldots
& {\pd f_{N-1}\over \pd \Gth_1}& f_{N-1} \cr
\vdots & & & \vdots\cr
{\pd^{N-1} f_0\over \pd \Gth_1^{N-1}}&\ldots
& {\pd f_0\over \pd \Gth_1} & f_0 \cr
}\right\vert\ ,}
where
\eqnn\IIIj
$$\eqalignno{f_{N-1}(\Gthd)&=\sum c_n^{(0)} p_{3n+N-1}(\Gthd) \cr
f_1(\Gthd)&=\sum c_n^{(1)} p_{3n+N-2}(\Gthd) \cr
\vdots & & \IIIj \cr
f_0(\Gthd)&=\sum c_n^{(N-1)} p_{3n}(\Gthd)\ . \cr
}$$
In the above we have used eq.~\IIad\ which is
meaningful for Schur functions with
independent $\Gth$ arguments. The precise meaning of \IIIj\ for
unconstrained $\Gth$'s is therefore that it extracts from the complete
formula \IIbc\ only those terms corresponding to Young  tableaux
which have at most $N$ rows. By letting $N \to \infty$ we reach
any desired term.

The expression \IIIi\ singles out the variable $\Gth_1$, the others
playing the role of parameters. For the time being we simplify the
notations by referring to $\Z$ as $Z$ until at the end we restore
the correct subscript. Looking at \IIIi\ we see that it takes
the form of a Wronskian of the components $f_0,\ldots,f_{N-1}$
of a vector denoted
\def\uf{\underline{f}}
$\uf$. It is therefore
natural to attach it to an $N$--th order differential operator
$\GD_N$ such that for an arbitrary function $F$ of $\Gth_1$
($\d\equiv {d\over d\Gth_1}$)
\eqn\IIIk
{\GD_N F = \sum_{r=0}^N w_r(\Gthd)\, \d ^{N-r} F
= Z^{-1} \left\vert \matrix{
{\pd^N F\over \pd\Gth_1^N} & \ldots & F \cr
{\pd^{N} f_{N-1}\over \pd \Gth_1^{N}}&\ldots
& f_{N-1}   \cr
\vdots & & \vdots\cr
{\pd^{N} f_0\over \pd \Gth_1^{N}}&\ldots
& f_0 \cr } \right\vert }
Strictly speaking the coefficients $w$ should also carry the subscript
$N$. In order to obtain a smooth transcription as $N\to \infty$, Sato
uses rather than the differential operator $\GD_N$ an equivalent
pseudodifferential operator $W$ defined as
\eqnn\IIIl
\eqnn\IIIm
$$\eqalignno{\GD_N&=W_N\, \d^N \ , &\IIIl \cr
W_N&=\sum_0^N w_r\, \d^{-r} &\IIIm \cr}$$
with $w_0=1$ and
\eqn\IIIn{w_1=Z^{-1}\, \big(-\dd{}{\Gth_1}\big)\, Z.}
Expanded in power series in the $\Gth$'s, $w_1$ will have terms of
fixed degree independent of $N$ for $N$ large enough and the remark
applies to the successive coefficients $w_p$ (infinitely many as
$N\to \infty$) justifying that we drop eventually all reference to $N$.
Equation \IIIn\ admits a rather neat generalization as
\eqn\IIIo{w_r={1\over Z}p_r\bigg(- {\pd\over \pd\Gthd}\bigg) Z\ ,}
where
\eqn\IIIp{{\pd \over \pd \Gthd}\equiv {\pd\over \pd\Gth_1},
\oh {\pd\over\pd\Gth_2},\ldots,{1\over k}{\pd\over\pd\Gth_k},\ldots}
(which is meaningful since we consider the $\Gth$'s as independent
variables). Indeed let us form the generating function
\eqn\IIIq{Z^{-1}\bigg(\sum_{r\ge 0} y^r p_r\Big(-{\pd\over \pd\Gthd}\Big)
\bigg)Z= Z^{-1} \bigg(\exp-\sum_1^{\infty}{y^r\over r}{\pd\over\pd\Gth_r}
\bigg) Z.}
Acting on any column vector of $Z$, the operator ${\pd\over\pd\Gth_r}$
is equivalent to ${\pd^r\over\pd\Gth_1^r}$. This gives
\eqn\IIIr{
Z^{-1}\bigg(\sum_{r\ge 0} y^r p_r\Big(-{\pd\over \pd\Gthd}\Big)
\bigg)Z=
 Z^{-1} \det \Big\vert(1-y\d) \d^{N-1}\uf,\ldots,(1-y\d)\uf \Big\vert\ .}
Let us compare this with the quantity
\eqn\IIIs{\sum_0^N w_r y^r =Z^{-1}
\det\left\vert\matrix{y^0 &\ldots&y^N\cr
		    \d^N\uf& \ldots &\uf\cr}\right\vert\ .}
The $(N+1)\times(N+1)$ determinant on the right hand side can be
computed by subtracting the first column multiplied by $y$ from the
second, the second column multipled by $y$ from the third and so on,
with the cofactor of the only non-vanishing element in the first row
equal to the previous expression proving   eq. \IIIo.

One is familiar with the commutation relations involving the operators
$\d^{-r}$
\eqnn\IIIt
$$\eqalignno{\d^{-r}a &= \sum_{k\ge0} (-1)^k {r+k-1\choose k}
a^{(k)} \d^{-r-k} \cr
a \d^{-r}&=\sum_{r\ge 0} {r+k-1 \choose k} \d^{-r-k}a^{(k)} &\IIIt\cr }$$
where $a^{(k)} \equiv (\d ^k a)$. These formulae enable one to give a
meaning, again droping the index $N$, to the ``dual'' pseudo--differential
operator
\eqn\IIIu{W^*=\sum \d^{-r} w^*_r \qquad\qquad w_r^*=Z^{-1} p_r
\bigg({\pd \over \pd \Gthd}\bigg) Z}
satisfying
\eqn\IIIv{W^*=W^{-1}}
We relegate the (cumbersome) proof of this latter fact
which relies on \P formulae to an appendix of this section.

The crux of the matter is the basic equation
\eqn\IIIw{{\pd W\over\pd\Gth_n}= Q_n W -W\d^n}
with $Q_n$ a normalized differential operator of order $n$,
$Q_n=\d^n+\ldots$, given by
\eqn\IIIx{Q_n=\bigg(W\d^n W^{-1}\bigg)_+}
the subscript $+$ refers to the differential part
of $W\d^n W^{-1}$, as follows from the
fact that ${\pd W\over \pd\Gth_n}$ is of order $\d^{-1}$.

Multiplying it on the right by $\d^N$, eq. \IIIw\  is equivalent to
\eqn\IIIy{{\pd \GD_N\over\pd\Gth_n}=Q_n \GD_N -\GD_N \d^n
\qquad \qquad Q_n=\bigg(\GD_N \d^n\GD_N^{-1}\bigg)_+}
where $\GD_N^{-1}=\d^{-N} W^{-1}$ . Again $N$ is assumed much larger
than $n$ fixed, in fact as large as we want. Among these equations
one is trivially verified namely the one for $n=1$ where $Q_1=\d$. To
prove \IIIy\  choose $Q_n$ as indicated so that the combination
$Q_n \GD_N-\GD_N\d^n$ is a differential operator of order $N-1$ as it
should be if the formula is to make sense. It will then hold if both sides
agree when operating on $N$ linearly independent functions which
(see eq. \IIIk) we naturally choose as being $f_0,\ldots,f_{N-1}$ or
equivalently on any linear combination which we denote by $f$. Thus
$\GD_N f=0$ and we want to prove that
\eqn\IIIz{\Big({\pd\GD_N\over\pd\Gth_n}\Big)f +\GD_N {d^n f\over d\Gth_1^n}
=0.}
But ${d^n f\over d\Gth_1^n}$ through the definition \IIIj\ is equal to
${}\pd f\over \pd \Gth_n$ reducing the above expression to
${\pd\over\pd\Gth_n}\big( \GD_N f\big)=0$. Thus eqs.~\IIIy\ and \IIIx\
hold true and yield in general the so-called KP hierarchy of compatible
integrable systems $ {\pd^2W\over\pd\Gth_{n_1}\pd\Gth_{n_2}}=$
$ {\pd^2W\over\pd\Gth_{n_2}\pd\Gth_{n_1}}$ for the coefficients of the
pseudo-differential operator $W$. An equivalent form is as follows. Set
\eqnn\IIIaa
$$\eqalignno{L&= W \d W^{-1} = \d + {\rm O}(\d^{-1}) \cr
L^n&= W \d^nW^{-1} \qquad\qquad Q_n=\big(L^n\big)_+\ . &\IIIaa \cr }$$
Then from  \IIIw\ we have
\eqnn\IIIab
$$\eqalignno{{\pd W^{-1}\over \pd\Gth_n}=& -W^{-1}Q_n +\d^n W^{-1}\cr
{\pd L\over\pd\Gth_n}=& [Q_n,L]\ , &\IIIab \cr}$$
where in writing these formulae we have implicitly taken the $N\to \infty$
limit. As a consequence of \IIIab\ we have the zero curvature conditions
\eqn\IIIac{{\pd Q_m\over\pd\Gth_n}-{\pd Q_n\over \pd\Gth_m}+[Q_m,Q_n]=0.}
According to equations \IIIo\ and \IIIu\  the coefficients in $W$
and $W^{-1}$ do not depend on
$\Gth_{2r}$. Hence  the differential operators
$Q_{2r}\equiv \big(L^{2r}\big)_+$ commute with $L$ as well as any of its
powers
\eqn\IIIad{[Q_{2r},Q_{2r'}]=0.}
Using the notations of theorem 1, $t_r=-(2r+1)!!\, \Gth_{2r+1}$,
$u={\pd^2\over\pd\Gth_1^2}\ln Z={\pd^2\over\pd t_0^2}\ln Z$ and \IIIo,
\IIIu\ and \IIIx
\eqnn\IIIae
$$\eqalignno{Q_2=& \d^2+2 u \cr
Q_3=& \d^3+3 u\d +{3\over 2} {\pd u\over\pd\Gth_1}\equiv
\big(Q_2^{{3\over 2}}\big)_+ & \IIIae \cr}$$
Setting $m=2$ and $n=3$ in \IIIac\ we conclude that the first non trivial
equation in the hierarchy reads
\eqn\IIIaf{{\pd Q_2\over\pd\Gth_3}=[Q_3,Q_2]}

\it i.e. \rm
\eqn\IIIag{{\pd u\over\pd t_1}={\pd\over\pd t_0}\bigg({1\over 12}{\pd^2 u
\over \pd t_0^2 } +\oh u^2 \bigg)}
as claimed in the second part of Theorem 1. Higher equations involve
${\pd\over\pd t_{2}}, \cdots$ and are of the form \IIhj{b},\IIhk.

In fact, the commutation of $L$ and $Q_2$ implies that $L^2=Q_2$, \it
i.e.\ \rm that $L^2$ is a differential operator.  To prove this, one may
appeal to a lemma \DS\ that asserts that the space of operators that
commute with $Q_2$ is spanned by the powers of the pseudodifferential
operator square root of $Q_2$, with {\it constant} coefficients.
Thus
\eqn\IIIaga{L=Q_2^{\oh}+\sum_{l=0}^{\infty}\Ga_l \(Q_2^{\oh}\)^{-l}\ . }
Both $L$ and $Q_2^{\oh}$, however, are functionals of $Z$ with the
limit $\d$ as $Z\to 1$. It follows that all the constants $\Ga$ vanish
and
\eqn\IIIagb{L=Q_2^{\oh}.}
This implies that the KdV flows \IIIab\ are generated by the
  $$Q_{2r+1}= (L^{2r+1})_+= (Q_2^{r+\oh})_+ . $$

Notice that the above identity \IIIagb\ means that $L$, which \it
a priori \rm  depends on all the derivatives of $Z$, is actually a
functional of the sole $u=\pd^2 \ln Z / \pd \Gth_1^2$.

\bigskip
%
{\it Appendix}\par
\noindent
It would seem that \P relations have not entered directly the discussion.
One place where they play a hidden role is in the computation of the
inverse $W^{-1}=W^*$. Of course if we need only the first few terms as
in \IIIae\ one can obtain them by a direct calculation. For completeness
we give a recursive proof of eq.~\IIIv. The integer $N$ being fixed we
start with the expressions \IIIi--\IIIk\ and consider $\uf$ in
\IIIj\ as a column vector function of independent $\Gth$'s with $\d\equiv
{\pd\over\pd\Gth_1}$ and $\uf^{(r)}\equiv {\pd\uf\over\pd\Gth_1^{r}}$.
Dropping the index $N$
\eqnn\IIIah
$$\eqalignno{\GD=& W \d^N \cr
W=&\sum_0^N w_r \d^{-r}&\IIIah \cr
w_r=& Z^{-1} p_r\bigg(-{\pd\over\pd\Gthd}\bigg)Z=Z^{-1}(-1)^r
\Big\vert\uf^{(N)}\ldots\widehat{\uf^{(N-r)}}\ldots \uf\Big\vert \cr }$$
using a shorthand notation for determinants. The kernel of $\GD$ is
the finite dimensional vector space generated by the components of
$\uf$. We distinguish a flag $(f_0),(f_0,f_1),(f_0,f_1,f_2), \cdots$
and associate to it a sequence of determinants
\eqn\IIIai{ Z^{(1)}=f_0 \qquad Z^{(2)}=\left\vert\matrix{f_1' &f_1\cr
                                                 f_0'&f_0\cr}
\right\vert, \cdots,  Z^{(N)}\equiv Z}
in terms of which one can write a factorized form  (the
Miura transformation)
\eqn\IIIaj{\GD=\Big({ Z^{(N)}\over
Z^{(N-1)}}\d {Z^{(N-1)}\over  Z^{(N)}}\Big)
\cdots \Big({
 Z^{(2)}\over Z^{(1)}}\d { Z^{(1)}\over  Z^{(2)}}\Big)
\Big( Z^{(1)}\d {1\over  Z^{(1)}}\Big) \ .}
It is clear that applied to $ Z^{(1)}=f_0$, $\GD$ gives $0$ while if
$\GD_k$ is the product of the first $k$ factors starting from the
right and if we assume $\GD_k f_0=\GD_k f_1=\ldots=\GD_k f_{k-1}=0$, then
$\GD_k f_k={Z^{(k+1)} \over  Z^{(k)}}$, hence
$$\GD_{k+1}f_k=\Big( {Z^{(k+1)}\over Z^{(k)}}\,\d\, {Z^{(k)}
\over  Z^{(k+1)}}\Big) f_k=0$$
proving the above factorization.
The identity to be established is therefore
\eqnn\IIIak
$$\eqalignno{
W^{-1}&=\d^N \GD^{-1}=\d^N \Big( Z^{(1)}\dm {1\over  Z^{(1)} }\Big)
\Big({ Z^{(2)}\over Z^{(1)}}\dm { Z^{(1)}\over  Z^{(2)}}\Big)\cdots
\Big({ Z^{(N)}\over Z^{(N-1)}}\dm {Z^{(N-1)}\over  Z^{(N)}}\Big)\cr
&= \sum_{r\ge 0} \d^{-r} w_r^* &\IIIak\cr }$$
with coefficients $w_r^*$ given by
\eqn\IIIal
{w_r^*=Z^{-1} p_r\bigg({\pd\over\pd\Gthd}\bigg)Z\ . }
Upon taking  a generating function
\eqn\IIIam{\sum_{r\ge 0} y^rw_r^*=Z^{-1} \exp \Big(\sum_1^{\infty}
{y^n\over n} {\pd \over\pd \Gth_n}\Big)Z= Z^{-1} \left\vert
{1\over 1-y{\pd\over\pd \Gth_1}} \uf^{(N-1)}\ldots
{1\over 1-y{\pd\over\pd \Gth_1}} \uf \right\vert }
where upper indices on $\uf$ label derivatives.
We have recognized that acting on each column of $Z$ the shift operator
is equivalent to
$$\exp \sum_1^{\infty} {y^n\over n}\left(
{\pd\over\pd\Gth_1}\right)^n={1\over 1-y{\pd\over\pd\Gth_1}}.$$
If in the above determinant we subtract from
the last column the preceding one multiplied by $y$
and so on, we get
\eqn\IIIan{w_r^*=Z^{-1}\vert\uf^{(N-1+r)}, \uf^{(N-2)}\cdots \uf\vert.}
Comparing \IIIah\ and \IIIak, we see that the determinental numerators
have a natural pictorial description in terms of Young tableaux. The first
determinant is a vertical Young tableau,
the second a horizontal one. This parallels the correspondence
between $p_r(-\Gthd)=(-1)^r s_r(\Gthd)$ (recall \IIaa{}) and $p_r(\Gth)$,
and the formula to be established is similar to the identity
$\det(\un -X) \det(\un-X)^{-1}=1$.

In any case with this expression
for $w_r^*$ we return to \IIIak\ and note that \IIIan\ says that $w_0^*=1$
in agreement with \IIIak\ for every $N$ whereas, should $N=1$, $W^{-1}$
reduces to
\eqn\IIIao{\d\big(f_0\dm f_0^{-1}\big)=\sum_{r\ge 0}\d^{-r}{f_0^{(r)}\over
f_0}}
again in agreement with \IIIan. We therefore assume that \IIIan\ holds for
any $r$ if $N' < N$ and for $r'\le r$ when the size of determinants
is $N$, and establish it for $w_{N,r+1}^*$ reinstating the index $N$. We have
\eqnn\IIIap
$$\eqalignno{W_N^{-1}&=\d W_{N-1}^{-1} { Z^{(N)}\over Z^{(N-1)}}\dm{Z^{(N-1)}
\over Z^{(N)} }\cr
&= \sum_{k\ge 0}\d ^{1-k}w_{N-1,k}^*
 { Z^{(N)}\over Z^{(N-1)}} \dm { Z^{(N-1)} \over  Z^{(N)} }& \IIIap \cr
&= \sum_{r\ge 0}\d^{-r} \sum_{k+l=r} \Big(w_{N-1,k}^* { Z^{(N)}\over
Z^{(N-1)} }\Big)^{(l)} {Z^{(N-1)}\over  Z^{(N)}}. \cr}$$
Write
\eqn\IIIaq{w_{N,r}^*={v_{N,r}\over  Z^{(N)}}.}
We have
\eqn\IIIar{v_{N,r}=Z^{(N-1)} \sum_{k+l=r} \left(v_{N-1,k}{ Z^{(N)}\over
Z^{(N-1)\,2}}
\right)^{(l)}\ .}
We want to show that
\eqn\IIIas{v_{N,\Gr}=\vert\uf^{(N-1+\Gr)}, \uf^{(N-2)}\ldots \uf\vert}
assuming it to be true for $N'<N$ (where we only keep the
components $f_0,\ldots,f_{N'-1}$) and also for $\Gr\le r$ to prove that it
holds for $r+1$. Take a derivative of the above identity
\eqn\IIIat{v_{N,r}'= {Z^{(N-1)}{}'\over Z^{(N-1)}} v_{N,r}
+Z^{(N-1)} \sum_{k+l=r} \left(v_{N-1,k}{ Z^{(N)}\over Z^{(N-1)\,2}}
\right)^{(l+1)}.}
The last sum differs from $v_{N,r+1}$ by the missing term
$v_{N-1,r+1} { Z^{(N)}\over Z^{(N-1)}}$. Hence
\eqn\IIIau{v_{N,r+1}=v'_{N,r}+ v_{N-1,r+1} {Z^{(N)}\over Z^{(N-1)}}
{Z^{(N-1)}{}' \over Z^{(N-1)}} v_{N,r} .}
\def\uph{\underline{\varphi}}
Denote the column vector $\big(f_0,\cdots, f_{N-2}\big)^T$ by $\uph$
(of dimension $N-1$). According to  the recursive hypothesis this reads
\eqn\IIIav{v_{N,r+1}=\vert \uf^{(N+r)},\uf^{(N-2)},\cdots,\uf\vert+{\Gc\over
Z^{(N-1)}} }
with
\eqnn\IIIaw
$$\eqalignno{
\Gc&=\Big\vert \widehat{\uph^{(N-1+r}},\widehat{\uph^{(N-1)}},\uph^{(N-2)}
\ldots,\uph \Big\vert
\Big\vert \uf^{(N-1+r)},\uf^{(N-1)},\widehat{\uf^{(N-2)}},
\cdots,\uf\Big\vert \cr
&- \Big\vert \widehat{\uph^{(N-1+r}},{\uph^{(N-1)}},\widehat{\uph^{(N-2)}}
\ldots,\uph \Big\vert
\Big\vert \uf^{(N-1+r)},\widehat{\uf^{(N-1)}},{\uf^{(N-2)}},
\cdots,\uf\Big\vert \cr
&+ \Big\vert {\uph^{(N-1+r}},\widehat{\uph^{(N-1)}},\widehat{\uph^{(N-2)}}
\ldots,\uph \Big\vert
\Big\vert\widehat{ \uf^{(N-1+r)}},\uf^{(N-1)},{\uf^{(N-2)}},
\cdots,\uf\Big\vert  &\IIIaw\cr}$$
where in each term the first determinant is $(N-1)\times (N-1)$ dimensional,
the second $N\times N$. We have to show that the
combination $\Gc$ vanishes since we wish to prove that $v_{N,r+1}$ is
the first term of the r.h.s. of \IIIav.
One easily checks that $\Gc=0$ if $N=2$,
so we henceforth assume $N>2$. To reduce the vanishing of $\Gc$ to one
of \P's identities, expand the $N\times N$ determinants involving $f$ and
its derivatives according to its first line. For each term of the form
$f_{N-1}^{(k)}$ the coefficient is a combination of $\uph$--determinants
which vanishes by virtue of the \P relations \IIIh, completing the proof of
formulas \IIIu\ and \IIIv.

%
%
\newsec{Matrix Airy equation and Virasoro highest weight conditions.}
\noindent The differential equation \IIkk\ generalizes to the
$N$--dimensional case as follows. Call $Y(\GL)$ the integral
appearing in \IIl\ for finite $N$
\eqn\IVa{Y(\GL)= \int dM \, \exp i \tr\({M^3\over 6}+{M\GL^2\over 2}\).}
The function $Y$ satisfies for each index $k$
\eqn\IVb{0= \int dM {d \over dM_{kk}}
\exp{i\over 6} (3 \tr M \GL^2  + M^3 ) }
\it i.e.\rm
\eqn\IVc{
0= \Big<\sum_{l,\, l\ne k}M_{kl}M_{lk}+ M_{kk}^2+\Gl_k^2 \Big> }
where $\langle . \rangle$ denotes an integral taken with respect to
the weight $dM \exp i \tr \( {M^3\over 6 }+{M\GL^2\over 2}\)$.
The insertion of a diagonal factor $M_{kk}$ can be achieved by acting with
the derivative operator $-i {1\over \Gl_k}\dd{}{\Gl_k}$ on $Y$. To deal
with non-diagonal insertions we express the invariance of the integral $Y$
under an infinitesimal change of variable of the form
\eqn\IVcc{ M \to M+i\Ge [X,M], {\rm \quad with \quad }
X_{ab}=\Gd_{ak}\Gd_{bl}M_{kl}.}
The Jacobian is $1+i\Ge (M_{ll}-M_{kk})$,
while the term $\tr M^3$ is invariant. Thus
\eqn\IVd{0= \Big< M_{ll}-M_{kk}+{i\over 2} (\Gl_k^2-\Gl_l^2)M_{kl}M_{lk}
\Big>}
with no summation implied.
Inserting this into \IVc\ leads to\foot{
For an alternative derivation one can transform the ``equation of motion''
\IVc\ into a matrix differential equation, assuming at first the
argument $T\equiv \GL^2$ to be an arbitrary (\it i.e. \rm not necessarily
diagonal) Hermitian matrix. Recognizing that the integral is invariant
under conjugation of $T$, hence only a function of its eigenvalues
$\{t_a\}$ one then uses
$$ \dd{}{T_{kl}}=\sum_a\dd{t_a}{T_{kl}} \dd{}{t_a}=\sum_a
{{\rm Min}_{kl}(t_a-T)\over P'(t_a)}\dd{}{t_a} $$
where $P(x)=\det(x-T)$ and ${\rm Min}_{kl}$ denotes the $(k,l)$ minor in the
corresponding matrix.}
\eqn\IVe{0= \Gl_k^2 + \langle M_{kk}^2 \rangle -2i\sum_{l, l\ne k}
{\langle M_{kk}- M_{ll}\rangle \over \Gl_k^2-\Gl_l^2}.}
This yields the matrix Airy equations
\eqn\IVf{\Big[ \Gl_k^2 -\(\inv{\Gl_k}\dd{}{\Gl_k}\)^2
-2 \sum_{l, l\ne k}
{1\over \Gl_k^2-\Gl_l^2}\(\inv{\Gl_k}\dd{}{\Gl_k}-\inv{\Gl_l}\dd{}{\Gl_l}\)
\Big]Y=0}
which can be turned into equivalent equations for $Z$ itself
\eqnn\IVfa
$$\eqalignno{\[ \inv{\Gl_k^2}\(\sum_l \inv{\Gl_l}\)^2 \right.&
	+\inv{4\Gl_k^4}+2\sum_{l, l\ne k} \inv{\Gl_k^2-\Gl_l^2}
\(\inv{\Gl_k}\dd{}{\Gl_k}-\inv{\Gl_l}\dd{}{\Gl_l}\) \cr
\qquad &\left. -2\( 1+\inv{\Gl_k^2}\sum_l\inv{\Gl_k+\Gl_l}\)\dd{}{\Gl_k}
+\(\inv{\Gl_k}\dd{}{\Gl_k}\)^2\] Z=0 \ .& \IVfa \cr }$$
In the limit $N\to \infty$ we know from sec.~3 that $Z$ admits an expansion
in terms of odd traces
\eqn\IVg{t_n=-(2n-1)!! \sum_l \inv{\Gl_l^{2n+1}}.}
The differential equations \IVf\ can be expanded in
inverse powers of $\Gl_k$ in the form
\eqn\IVh{2\sum_{m\ge -1}\inv{(\Gl_k^2)^{m+2}}L_m Z=0.}
Explicitly
\eqnn\IVi
$$\eqalignno{
L_{-1}=& \oh t_0^2 +\sum_{k\ge 0} t_{k+1}\dd{}{t_k} -\dd{}{t_0} \cr
L_0=& \inv{8}+\sum_{k\ge 0} (2k+1)t_k \dd{}{t_k} -3\dd{}{t_1} &\IVi\cr
L_1=& \sum_{k\ge 1}(2k+1)(2k-1)t_{k-1}\dd{}{t_k} +\oh\dd{{}^2}{t_0^2}
-15\dd{}{t_2} \cr
L_2=& \sum_{k\ge 2}(2k+1)(2k-1)(2k-3)t_{k-2}\dd{}{t_k} +3{\pd^2\over
\pd t_0\pd t_1}-105\dd{}{t_3} \cr
& \ldots\ldots\ldots \cr
L_m=& \sum_{k\ge m}{(2k+1)!!\over (2(k-m)-1)!!}t_{k-m}\dd{}{t_k}
+\oh\sum_{k+l=m-1} (2k+1)!!(2l+1)!!{\pd^2\over\pd t_k\pd t_l}\cr
&\qquad -(2m+3)!! \dd{}{t_{m+1}}
 +{t_0^2\over 2}\Gd_{m+1,0}+\inv{8}\Gd_{m,0}\cr
}$$
\medskip
\noindent In eq.~\IVh, each coefficient has to vanish so that\par
\smallskip\penalty -300
\noindent \bf Theorem 2\rm \ (\K)\par \penalty 600
\noindent \it $Z$ satisfies and is determined by the highest weight
conditions
\eqn\IVj{L_m Z=0 \qquad\qquad m\ge -1.}
\rm
\smallskip
The operators $L_m$ obey (part of) the Virasoro (or rather the Witt)
algebra, namely
\eqn\IVk{[L_m,L_n]=(m-n) L_{m+n}\qquad\qquad m,n\ge -1 }
generated by $L_{-1}, L_0,L_1,L_2$. Note that only the first two
involve first order derivatives.

\newsec{Genus expansion.}
\noindent
As in standard matrix models there exists a genus expansion for $F$
\eqn\Va{\ln Z = F= \sum_{g\ge 0}F_g}
One way to obtain it is from the Airy system.
One inserts appropriate extra factors of $N$ and studies the
large $N$ limit, paying attention to corrections, according to a method
applied to leading order by Kazakov and Kostov \KK\ and revived by Makeenko
and Semenoff \MS. In the spirit of our paper we
follow a slightly different approach based on the KdV equations and
Virasoro constraints. In genus $g$, $F_g$ collects in the
expansion \IIa\ all terms such that
\eqn\Vb{F_g(\tdot)=\sum_{{k_i\atop \Sigma_{i\ge 0}(i-1)k_i=3g-3}} \left<
\prod_{i\ge 0}{(\tau_i t_i)^{k_i}\over k_i!}\right>}
We set
\eqn\Vc{u_g(\tdot)={\pd^2 F_g\over \pd t_0^2}\ .}
In the KdV hierarchy (eq.~\IIhj{}) the leading term in the semi-classical
or genus expansion corresponding to the term $u_0$ is obtained
from power counting by ignoring in the differential polynomial $R_n$
all terms involving derivatives. It then reduces to
\eqn\Vd{n\ge 0\qquad\qquad \dd{u_0}{t_n}=\dd{}{t_0}{u_0^{n+1}
\over (n+1)!}\ .}
For $n=0$ this is vacuous and has to be supplemented
by the first Virasoro condition \IVh\ (for $m=-1$) which amounts to
\eqn\Ve{\dd{u_0}{t_0}=1+\sum_{k\ge 0}t_{k+1}\dd{u_0}{t_k}\ .}
Inserting \Vd\ into \Ve\ gives
\eqn\Vf{\dd{}{t_0}\left\{u_0-\sum_{n\ge 0}t_n{u_0^n\over n!}\right\}=0\ .}
Define
\eqn\Vg{I_k(u_0,\tdot)= \sum_{p\ge 0}t_{k+p}{u_0^p\over p!}\ .}
Equation \Vf\ suggests
\par \medskip
\noindent \bf Lemma 4 \rm \par
\noindent \it $u_0(\tdot)$ satisfies the implicit equation
\eqn\Vh{u_0-I_0(u_0,\tdot)=0.}
\rm
\smallskip
\noindent To check this, multiply \Ve\ by ${u_0^n\over n!}$ and use \Vd\
to obtain
\eqn\Vi{\dd{u_0}{t_n}={u_0^n\over n!}+\sum_{k\ge 0}t_{k+1}\dd{}{t_k}
{u_0^{n+1}\over (n+1)!}}
while
\eqn\Vj{\dd{I_0(u_0,\tdot)}{t_n}={u_0^n\over n!}+\sum_{k\ge 0}
t_{k+1}\dd{}{t_n}{u_0^{k+1}\over (k+1)!}\ .}
Subtracting and using \Vd\ again, we find
\eqnn\Vk
$$\eqalignno{\dd{}{t_n}\BBL u_0-I_0(u_0,\tdot)\BBR= &
\sum_{k\ge 0}t_{k+1}\left( {u_0^n\over n!}\dd{}{t_k}u_0-{u_0^k\over k!}
\dd{}{t_n}u_0 \right) & \Vk\cr
=& \sum_{k\ge 0}t_{k+1}\left({u_0^n\over n!}\dd{}{t_0}{u_0^{k+1}\over (k+1)!}
-{u_0^k\over k!}\dd{}{t_0}{u_0^{n+1}\over (n+1)!}\right) =0 \ .\cr }$$
The difference $u_0-I_0(u_0,\tdot)$ is thus a  constant. Since it vanishes
at $\tdot=0$ it is identically zero, completing the proof. To find $F_0$
we have to integrate twice eq.~\Vh\ with respect to $t_0$. From \Vb\ the
boundary conditions are given by the vanishing of $F_0$ and
$\dd{}{t_0} F_0$ when
$t_0=0$. This yields with $u_0(\tdot)$ implicitly given by \Vh\
\eqn\Vl{F_0= {u_0^3\over 6}-\sum_{k\ge 0}{u_0^{k+2}\over k+2}{t_k\over k!}
+\oh\sum_{k\ge 0}{u_0^{k+1}\over k+1}\sum_{a+b=k}{t_a\over a!}{t_b\over b!}\ .
}
\medskip
\noindent\bf Remarks \rm\par
\noindent (i) As is generally  the case if we extend $F_0$ by considering
$u_0$ (originally equal to $\pd^2 F_0/ \pd t_0^2$) as an auxiliary
\it independent \rm parameter, we find that the stationarity condition
\eqn\Vm{\dd{F_0}{u_0}=\oh \( u_0-I_0(u_0,\tdot)\)^2=0 }
yields equation \Vh. \par
\noindent (ii) The expression for $F_0$ is equivalent to the one given
by Makeenko and Semenoff \MS\ using (infinitely many) eigenvalues
$\Gl_k$
\eqnn\Vn
$$\eqalignno{F_0=&\inv{3} \sum_k \Gl_k^3-\inv{3}\sum_k (\Gl_k-2s)^{{3\over 2}}
-s\sum_k (\Gl_k^2-2s)^{\oh}\cr
& \qquad + {s^3\over 6} -\oh \sum_{k,l}\ln {\sqrt{\Gl_k^2-2s}+\sqrt{\Gl_l^2
-2s}\over \Gl_k+\Gl_l} & \Vn \cr}$$
with the condition
\eqn\Vo{\dd{F_0}{s}=\oh\(s+\sum_k\inv{\sqrt{\Gl_k^2-2s}}\)=0}
upon identification of $s$ with $u_0$, of $I_p$ with
 $-(2p-1)!! \sum_k \inv{(\Gl_k^2-2s)^{p+\oh}}$
and $t_n$ is  as in \IVg.
\par
\noindent (iii) From \Vl\ or \Vh\ we can readily find the first few terms
in the expansion of $F_0$ which up to a factor $t_0^3$ only involves the
combinations $t_0^{k-1} t_k$.
\eqnn\Vp
$$\eqalignno{
F_0&=\frac{t_0^3}{3!} + t_1 \frac{t_0^3}{3!}  + \(t_2 \frac{t_0^4}{4!}
+ 2 \frac{t_1^2}{2!} \frac{t_0^3}{3!}\)
 + \(t_3 \frac{t_0^5}{5!} + 3 t_1 t_2 \frac{t_0^4}{4!} + 6 \frac{t_0^3}{3!}
\frac{t_1^3}{3!}\)\cr &
+ \[t_4 \frac{t_0^6}{6!} +\(6 \frac{t_2^2}{2!} + 4 t_1 t_3\)
\frac{t_0^5}{5!} + 24 \frac{t_0^3}{3!} \frac{t_1^4}{4!} +
12 t_2 \frac{t_1^2}{2!} \frac{t_0^4}{4!}\]\cr
&  +\[t_5 \frac{t_0^7}{7!}+\(5 t_1 t_4 + 10 t_2 t_3\)\frac{t_0^6}{6!}
+ 120 \frac{t_0^3}{3!} \frac{t_1^5}{5!} +\(30 t_1 \frac{t_2^2}{2!}
 + 20 t_3 \frac{t_1^2}{2!}\)\frac{t_0^5}{5!} + 60 t_2 \frac{t_1^3}{3!}
\frac{t_0^4}{4!}\]\cr &
 +\[t_6 \frac{t_0^8}{8!} +\(20 \frac{t_3^2}{2!} + 6 t_1 t_5 + 15 t_2 t_4\)
\frac{t_0^7}{7!} + 720 \frac{t_0^3}{3!} \frac{t_1^6}{6!}
+\(90 \frac{t_2^3}{3!} + 30 t_4 \frac{t_1^2}{2!} + 60 t_1 t_2 t_3\)
\frac{t_0^6}{6!} \right.\cr
&\qquad\left. +\(120 t_3 \frac{t_1^3}{3!} + 180 \frac{t_1^2}{2!}
\frac{t_2^2}{2!}\)\frac{t_0^5}{5!} + 360 t_2 \frac{t_0^4}{4!}
\frac{t_1^4}{4!}\]  + \ldots & \Vp\cr}$$
In genus zero all coefficients are positive integers (as opposed to
fractional) due to the smoother structure of $\CM_{0,n}$, ($n\ge 3$).
Indeed we have the obvious\par
\smallskip
\noindent\bf Lemma 5\rm\par
\noindent\it The class of formal power series in $t_0, t_1, \ldots$
which vanish at $\tdot=0$ with non-negative integral derivatives at the
origin is stable under\par
(i) addition \par
(ii) product \par
(iii) composition \par
\rm \noindent
To apply this to $u_0$ (and hence to $F_0$) we remark that the
sequence
\eqn\Vpa{ f_0=t_0, \qquad f_n=\sum_0^{\infty} f_{n-k}^k {t_k\over k!}}
has each of its derivatives at the origin which stabilizes to the
corresponding one of $u_0$ after finitely many steps.

To obtain the next terms we split the Virasoro constraints expressed on
$\ln Z$ as follows
\eqnn\Vq
$$\eqalignno{m&=-1  \qquad \cr
& {t_0^2\over 2} \Gd_{g,o}+ \sum_{k\ge 0} t_{k+1}
\dd{F_g}{t_k}-\dd{F_g}{t_0}=0 \cr
m&=0 \qquad\cr
& \inv{8} \Gd_{g,1}+\sum_{k\ge 0} (2k+1) t_k\dd{F_g}{t_k}-3\dd
{F_g}{t_1}=0 & \Vq \cr
m&=1 \qquad\cr
 & \sum_{k\ge 1} (2k+1)(2k-1)t_{k-1}\dd{F_g}{t_k}-15 \dd{F_g}{t_2}
+\oh{\pd^2F_{g-1}\over \pd t_0^2}+\oh\sum_{g_1+g_2=g}\dd{F_{g_1}}{t_0}
\dd{F_{g_2}}{t_0}=0 \cr
m&=2 \qquad \cr
& \sum_{k\ge 2}(2k+1)(2k-1)(2k-3) t_{k-2}\dd{F_g}{t_k}-105
\dd{F_g}{t_3}+3 {\pd^2 F_{g-1}\over \pd t_0\pd t_1}+3\sum
_{g_1+g_2=g}\dd{F_{g_1}}{t_0}\dd{F_{g_2}}{t_1}=0 \ ,\cr   }$$
while a similar splitting of the KdV equation \IIhj{} yields
\eqn\Vr{\dd{u_g}{t_1}= \dd{}{t_0}\( \inv{12} {\pd^2 u_{g-1}\over \pd t_0^2}
+\oh \sum_{g_1+g_2=g} u_{g_1}u_{g_2}\).}
For genus one this equation reads
\eqn\Vs{\dd{}{t_0}\left\{ \( \dd{}{t_1}-u_0\dd{}{t_0}\)\dd{F_1}{t_0}
-\inv{12} {\pd^2 u_0\over \pd t_0^2}\right\}=0.}
We have
\eqnn\Vt
$$\eqalignno{\dd{u_0}{t_0}={1\over 1-I_1}\qquad, & \qquad \dd{u_0}{t_1}=
{u_0\over 1-I_1}\cr
p\ge 1 \qquad \dd{I_p}{t_0}={I_{p+1}\over 1-I_1} \qquad, & \qquad
\(\dd{}{t_1} -u_0 \dd{}{t_0}\) I_p= \Gd_{p,1}  }$$
hence we can rewrite
\eqnn\Vu
$$\eqalignno
{\inv{12}{\pd ^2 u_0\over \pd t_0^2 }=& \inv{12}{I_2\over (1-I_1)^3}
= \inv{24}\dd{}{I_1} {I_2\over (1-I_1)^2}\cr
=& \dd{}{I_1}\dd{}{t_0}\( \inv{24}\ln \inv{1-I_1}\). & \Vu \cr}$$
If $F_1$ is a function of $t_.$ only through $I_1,I_2,  \ldots$, we
can rewrite
\eqn\Vv{\(\dd{}{t_1}-u_0\dd{}{t_0}\)\dd{}{t_0}F_1= \dd{}{I_1}\dd{}{t_0}F_1}
so that equation \Vs\ becomes
\eqn\Vw{\dd{}{t_0}\dd{}{I_1}\dd{}{t_0}\left\{F_1-\inv{24}\ln
\inv{1-I_1}\right\}=0.}
This suggests that
\eqn\Vx{F_1=\inv{24}\ln \inv{1-I_1}\ ,}
in agreement with the above hypothesis so that eq.~\Vr\ is satisfied.
A straightforward computation shows that the Virasoro conditions are
satisfied, proving \Vx.
\smallskip
\noindent \bf Remark \rm\par
\noindent
It is not unexpected that the genus one (or ``one-loop'') result involves as
usual a logarithm. Expanding $F_1$
\eqnn\Vy
$$\eqalignno{
 24F_1& =
 t_1 +\(\frac{t_1^2}{2!} + t_0 t_2\)
 +\(2 \frac{t_1^3}{3!} + t_3 \frac{t_0^2}{2!} + 2 t_0 t_1 t_2\) \cr
&  +\(6 \frac{t_1^4}{4!} + t_4 \frac{t_0^3}{3!} + 4 \frac{t_0^2}{2!}
\frac{t_2^2}{2!} + 6 t_0 t_2 \frac{t_1^2}{2!} + 3 t_1 t_3
\frac{t_0^2}{2!}\)\cr
&  +\(24 \frac{t_1^5}{5!} + t_5 \frac{t_0^4}{4!} + 24 t_0 t_2
\frac{t_1^3}{3!}+\(4 t_1 t_4 + 7 t_2 t_3\)\frac{t_0^3}{3!}
+ 16 t_1 \frac{t_0^2}{2!} \frac{t_2^2}{2!}
+ 12 t_3 \frac{t_0^2}{2!} \frac{t_1^2}{2!}\)\cr
&  +\[120 \frac{t_1^6}{6!} + t_6 \frac{t_0^5}{5!} + 120 t_0 t_2
\frac{t_1^4}{4!} +\(14 \frac{t_3^2}{2!} + 5 t_1 t_5 + 11 t_2 t_4\)
\frac{t_0^4}{4!} + 48 \frac{t_0^3}{3!} \frac{t_2^3}{3!}
+ 60 t_3 \frac{t_0^2}{2!} \frac{t_1^3}{3!}\right.\cr
& \qquad \left.+\(20 t_4 \frac{t_1^2}{2!}
+ 35 t_1 t_2 t_3\)\frac{t_0^3}{3!} + 80 \frac{t_0^2}{2!}
\frac{t_1^2}{2!} \frac{t_2^2}{2!}\]\cr
&  +\[720 \frac{t_1^7}{7!} + t_7 \frac{t_0^6}{6!} + 720 t_0 t_2
\frac{t_1^5}{5!} +\(6 t_1 t_6 + 16 t_2 t_5 + 25 t_3 t_4\)
\frac{t_0^5}{5!} + 360 t_3 \frac{t_0^2}{2!} \frac{t_1^4}{4!}\right.\cr
& \qquad
+ \left.\(84 t_1 \frac{t_3^2}{2!} + 118 t_3 \frac{t_2^2}{2!} + 30 t_5
\frac{t_1^2}{2!} + 66 t_1 t_2 t_4\)\frac{t_0^4}{4!}\right.\cr
& \qquad \left. + 288 t_1 \frac{t_0^3}{3!} \frac{t_2^3}{3!}
+\(120 t_4 \frac{t_0^3}{3!} + 480 \frac{t_0^2}{2!} \frac{t_2^2}{2!}\)
\frac{t_1^3}{3!} + 210 t_2 t_3 \frac{t_1^2}{2!} \frac{t_0^3}{3!}\]
 + \ldots & \Vy \cr }$$
All intersection numbers are of the form $\inv{24}\times$ (a positive
integer) since $I_1$ and $-\ln (1-I_1)$ belong to the class of
functions referred to in Lemma 5.

For higher genus the Ansatz
\eqn\Vz{F_g=\sum_{\sum_{2\le k\le 3g-2} (k-1)l_k=3g-3}
\langle \tau_2^{l_2}\tau_3^{l_3}\ldots \tau_{3g-2}^{l_{3g-2}}\rangle
\inv{(1-I_1)^{2(g-1)+\sum l_p}} {I_2^{l_2}\over l_2!}{I_3^{l_3}\over l_3!}
\ldots {I_{3g-2}^{l_{3g-2}}\over l_{3g-2}!}, }
which is a finite sum of monomials in $I_k/ (1-I_1)^{{2k+1\over 3}}$, the
number of which is $p(3g-3)$ (with $p(n)$ the number of partitions of $n$),
is consistent with the KdV equation \Vr.
Inserted into the latter, it allows one to compute the coefficients with
the result

\eqn\Vaa{F_2=\inv{5760}\left[5{I_4\over (1-I_1)^3} +29{I_3 I_2\over
(1-I_1)^4 } +28{I_2^3\over (1-I_1)^5}\right]\ .}
Hence
\eqn\Vab{\vev{ \tau_4} =\inv{1152}\qquad \vev{\tau_2 \tau_3}={29\over 5760}
\qquad \vev{\tau_2^3}={7\over 240}}
in agreement with Witten \Wun. The other intersection numbers
can be derived by expanding \Vaa\ in $\tdot$.

For genus 3 we find
\eqnn\Vac
$$\eqalignno{
F_3&= \inv{2903040}\left[ 35{I_7\over (1-I_1)^5} +539{I_6I_2\over(1-I_1)^6}
+1006{I_5I_3\over(1-I_1)^6}+4284{I_5I_2^2\over(1-I_1)^7} \right. \cr
& \left.\quad +607{I_4^2\over(1-I_1)^6}+13452{I_4I_3I_2\over(1-I_1)^7}
+22260{I_4I_2^3\over(1-I_1)^8}+2915{I_3^3\over(1-I_1)^7}
 \right. \cr
& \left.\quad
+43050{I_3^2I_2^2\over(1-I_1)^8}+81060{I_3I_2^4\over(1-I_1)^9}
+34300{I_2^6\over(1-I_1)^{10}}   \right] \ ,
 & \Vac \cr}$$
which yields the table
%
\medskip
$$\vbox{\tabskip 0pt\offinterlineskip
\halign to 13cm{
\qquad\qquad\qquad
\qquad\strut\hfil$#$ & $#$\hfil&\hfil$#$&\hfil$#$ & $#$\hfil \cr
\bra \tau_7\ket =& \inv{82944}&&
        \bra \tau_4\tau_3\tau_2\ket=& {1121\over 241920}\cr
\noalign{\vskip 5pt}
\bra \tau_6\tau_2\ket= &{77\over 414720}& &
        \bra \tau_4\tau_2^3\ket =& {53\over 1152}\cr
\noalign{\vskip 5pt}
\bra \tau_5\tau_3\ket= & {503\over 1451520}&&
        \bra \tau_3^3\ket=& {583\over 96768}\cr
\noalign{\vskip 5pt}
\bra \tau_5\tau_2^2\ket=&{17\over 5760}&&
        \bra \tau_3^2\tau_2^2\ket =& {205\over 3456}\cr
\noalign{\vskip 5pt}
\bra \tau_4^2\ket=&{607\over 1451520}&&
       \bra \tau_3 \tau_2^4\ket =& {193\over 288}\cr
\noalign{\vskip 5pt}
		& & \bra \tau_2^6\ket={1225\over 144}&&  \cr
}}$$
\centerline{Table I}
\bigskip
There is no difficulty to pursue these computations as far as one wishes.

\smallskip
\noindent
\bf Remarks \rm \par
\noindent
(i) It follows from Lemma 5 that all series coefficients in \Vz\ reexpressed
in terms of
$\prod t_k^{l_k}/ l_k!$ are non-negative integers up to the finitely
many prefactors.
All intersection numbers of a fixed genus when written as irreducible
fractions have therefore
a lowest common multiplier (l.c.m.) $D_g$: $D_0=1$,
$D_1=24$, and for $g>1$, $D_g$ is the lowest common denominator of the
finitely many intersection numbers appearing in \Vz, when written as
irreducible fractions.
We conjecture that for $1<g'\le g$, the order of any automorphism group of an
algebraic curve of genus $g'$ (bounded by $84(g-1)$) divides
$D_g$. Thus $D_3= 2 903 040$ is divible by 168 (the order of the largest
automorphism group of a genus 3 curve) and by 48 (the same for genus 2).

\noindent (ii) The term in $F_g$, $g>1$ which has the highest power of
$(1-I_1)$ in the denominator has the form
$$ {\vev{\tau_2^{3g-3}}\over (3g-3)! } {I_2^{3g-3}\over (1-I_1)^{5g-5}}$$
In the next section we develop a formalism to resum these terms as well
as subleading ones akin to the ``double scaling limit'' of standard
matrix models \sextet\ (one recognizes the same string exponents and
the same ingredients).
At the other extreme the term with the lowest power of $1-I_1$
in the denominator is \Wun
\eqn\Vae{\vev{\tau_{3g-2}}{I_{3g-2}\over (1-I_1)^{2g-1}}
\qquad \qquad \vev{\tau_{3g-2}}=\inv{(24)^g g!}\ .}
The last equality (also valid for $g=1$)
follows from \Vr\ by keeping terms
with the lowest power of $(1-I_1)^{-1}$.
It implies that
$(24)^g g!$ divides $D_g$.

\newsec{Singular behaviour and Painlev\'e equation.}

\noindent
The expression \IIn\ exhibits a singular behaviour as $I_1 \to 0$.
In a first step, we can keep in the KdV equation
the dominant terms by considering
that
\eqn\VI{ I_2 \approx {\rm constant},\qquad\qquad I_k\approx 0
\ {\rm for\  k\ge 3.}}
This
is consistent with the derivatives of the $I$'s: $\pd I_k/\pd t_0
= I_{k+1}/(1-I_1)$ and
$ \pd I_k / \pd t_1=u_0 I_{k+1}/(1-I_1)$ for $k\ge 2$.
In this approximation, the genus $g$ contribution to the specific heat
$u_g= \pd^2 F_g / \pd t_0^2$ is of the form
\eqn\VIa{u_g= \pd^2 F_g / \pd t_0^2=
\alpha_g {I_2^{3(g-1)+2}\over (1-I_1)^{5(g-1)+4}}\ .}
We introduce the scaling variable
\eqn\VIb{z={(1-I_1)\over I_2^{3/5}}}
and separate the genus zero contribution by setting $u=u_0 + \tilde u$,
$\tilde u= \sum_{g\ge 1}u_g$.
The KdV equation \IIhj{a} then reads
\def\tu{\tilde u}
\eqnn\VIc
$$\eqalignno
{{\pd \tu\over \pd t_1}=&
\tu\big( {1\over 1-I_1}
+{\pd \tu\over \pd t_0 }\big)
+{1\over 4}{I_2^2\over (1-I_1)^5}
+{1\over 12}{\pd^3 \tu \over\pd t_0^3}
\cr =&
\tu\big( {1\over 1-I_1}
+{\pd \tu\over \pd I_1}\big)
+{1\over 4}{I_2^2\over (1-I_1)^5}
+{1\over 12} \big({I_2\over 1-I_1}{\pd \tu \over\pd I_1}\big)^3 \ .
& \VIc\cr }$$
A rescaling
\eqn\VId{\tu= I_2^{-2/5} \psi(z)}
leads to the equation
\eqn\VIe{{\pd\psi\over\pd z}
+{\psi\over z} \big[1-{\pd \psi\over \pd z} \big]
+{1\over 4z^5}
-{1\over 12}\big({1\over z}{\pd \over \pd z}\big)^3 \psi =0.}
This equation will be generalized below. In this particular case of
the behaviour \VI, one can transform it into the Painlev\'e equation:
we set $t=2^{-2/5}z^2$, $\psi(z)= z+2^{1/5}\phi(t)$ and find
\eqn\VIf{\inv{3} \phi''+\phi^2 -t=0 \ .}
which has the asymptotic expansion
\eqn\VIfa{\phi =\sum{\phi_g\over t^{{5\over 2} (g-1)+2}} \qquad,\qquad
\phi_0=-1\ ,}
where the successive terms satisfy
\eqn\VIg{\phi_{g+1}={25g^2-1\over 24}\phi_g+\oh\sum_{m=1}^g
\phi_{g+1-m}\phi_{m} \ .}
Hence in this regime
\eqnn\VIga
$$\eqalignno{\sum_{g\ge 2}F_g^{{\rm sing}}&= \sum_{g\ge 2} {\vev{
\tau_2^{3g-3}} \over (3g-3)!} {I_2^{3(g-1)}\over (1-I_1)^{5(g-1)} } \cr
\vev{\tau_2^{3g-3}}&={2^g (3g-3)!\over (5g-5)(5g-3)} \phi_g &\VIga\cr
}$$
\eqn\VIgb{g=2\quad \vev{\tau_2^3}={7\over 240}, \qquad
g=3 \quad \vev{\tau_2^6}={1225\over 144}, \qquad g=4 \quad \vev{\tau_2^9}=
{1816871\over 48}, \ldots\ .}

\bigskip

This discussion may be extended to the regime in which all (or a finite
number of) the $I$'s
are retained and  tend to zero according to the following scaling law
\eqnn\VIh
$$\eqalignno{z&={(1-I_1)\over I_2^{3/5}}\cr
v_q&= {I_q(1-I_1)^{q-2}\over I_2^{q-1}}\qquad q\ge 3& \VIh \cr
F^{{\rm sing}}_g&= z^{-5(g-1)}
\sum_{\GS_{2\le k\le 3g-2} (k-1)l_k=3g-3}
{ \langle \tau_2^{l_2}\tau_3^{l_3}\ldots \tau_{3g-2}^{l_{3g-2}}\rangle
} \prod_{q=3}^{3g-2} {v_q^{l_q}\over l_q!} \cr
}$$
The KdV equation is then rephrased as
\eqnn\VIi
$$\eqalignno{
\Big[z{\pd \over \pd z}&+\sum_{q\ge 3}(q-2)v_q{\pd\over \pd v_q}+1\Big]\psi=
& \VIi\cr
&= {1\over z}\psi\big(\GD+{2\over 5}v_3\big)\psi -{3+v_3\over 12z^4}
+{1\over
12z^5}[\GD-4-{8\over 5}v_3][\GD-2-{3\over 5}v_3][\GD+{2\over 5}v_3]\psi
\cr }$$
where the same change of function as in \VId\ has been carried out, and
$\GD$ denotes the differential operator
\eqn\VIj{\GD=(1+{3\over 5}v_3)z{\pd\over \pd z}+\sum_{q\ge 3}
\big(-v_{q+1}+(q-2)v_q+(q-1)v_q v_3\big){\pd\over \pd v_q}\ .}
The contribution to a given genus $g$ involves only a finite number of
terms in the sum: $q\le 3g-2$. Moreover the differential operators
in \VIi\ respect the grading in powers of $z$,
\def\vdot{v_{\textstyle .}}
$$ \psi_g(z,\vdot) = z^{-5(g-1)-4} \psi_g(\vdot) \ ,$$
thus determining the polynomials $\psi_g(\vdot)$
recursively. For instance, if only $z$ and $v\equiv v_3$
are retained, we have for $g>1$ (and $\pd_v\equiv \pd/\pd v$)
\eqnn\VIk
$$\eqalignno{\[5(g-1)+3-v\pd_v \] \psi_g(v)&= \sum_{g'=1}^{g-1}
\psi_{g-g'}(v) \[4+2v+(g'-1)(5+3v)-(1+2v)v\pd_v\] \psi_{g'}(v) \cr
& +{1\over 12}
\[8+4v+(g-2)(5+3v)-(1+2v)v\pd_v\] \times
& \VIk\cr
 \[6+3v+(g-2)(5+3v)-\right.&\left.(1+2v)v\pd_v\]
\[4+2v+(g-2)(5+3v)-(1+2v)v\pd_v\]
\psi_{g-1}(v) \ ,\cr }$$
which yields
\eqnn\VIl
$$\eqalignno{
24\, \psi_1(v)=&{2+v} \cr
1152\, \psi_2(v)=&{196+352v+109v^2}&\VIl \cr
82944\, \psi_3(v)=&{117600+362564 v+324660 v^2+84699 v^3+3043 v^4} \cr
3981312\,\psi_4(v)=& 1906157232+ 7865959024 v
+ 11212604992 v^2+ 6581090736 v^3 \cr
& \qquad\qquad\qquad
+ 1465796801 v^4 +83580341 v^5
.\cr}$$
{}From this one extracts the first intersection numbers of the form
$\langle \tau_2^{l_2}\tau_3^{l_3}\rangle $. One recovers for $g\le 3$
results obtained in \Vab\ or in Table I, and in genus $g=4$ the results
\eqnn\VIm
$$\eqalignno{
\langle \tau_2^9\rangle =& {1816871\over48} \cr
\langle \tau_2^7 \tau_3 \rangle =&{3326267\over 1728}\cr
\langle \tau_2^5 \tau_3^2\rangle=& {728465\over 6912}\cr
\langle \tau_2^3 \tau_3^3\rangle =& { 43201\over 6912}\cr
\langle \tau_2^7 \tau_3^4\rangle=& {134233\over 331776}\ .\cr
}$$
\penalty 1000
\centerline{Table II}
\par
We hope we have amply demonstrated the practical use of these
expansions.

\newsec{Generalization to higher degree potentials}
\noindent
The cubic potential in the integral \Oa\ may be generalized to a
potential of degree $p+1$,
as noticed by several authors \AvM,\Kde,\Wtr. The case $p=2$ is the one
discussed previously. Let us consider first the one-variable
integral analogous to \IIi, also denoted $z(\Gl)$.
The normalizations are adjusted in such a way as to make the quadratic
term positive definite, in order to have a well-defined asymptotic expansion
%
\eqn\VIIa{ z(\Gl)=  {\int_{-\infty} ^{\infty}
dm\, e^{{i^{p^2+1}\over 2(p+1)}
\[ \(m+(-i)^{p+1} \Gl)\)^{p+1}\]_{>{\rm lin}}}
\over \int_{-\infty}^{\infty} dm\, e^{-{p\over 4}m^2\Gl^{p-1}}  }\ , }
where the subscript ``$>$ lin''
denotes the sum of terms of degree $\ge 2$ in the
polynomial. By considerations similar to those of sect.~2.1, it is
easy to see that $z(\Gl)$ admits an asymptotic expansion in inverse powers
of $\Gl^{p+1}$
\eqn\VIIb{z(\Gl)= \sum_0^{\infty} c_k \Gl^{-(p+1)k}}
with $c_0=1$. It satisfies a differential equation of order $p+1$
\eqn\VIIc{(D^p-\Gl^p) z(\Gl)=0}
\eqnn\VIId
$$\eqalignno{D=& \Gl^{{p-1\over 2}} e^{- {p\over 2(p+1)} (-\Gl)^{p+1}}
\(2(-1)^{p+1} \dd{}{\Gl^p} \) \Gl^{-{p-1\over 2}} e^{
 {p\over 2(p+1)} (-\Gl)^{p+1}}
\cr
=& \Gl+ {(-1)^p\over p}\( {p-1\over\Gl^p} - {2\over \Gl^{p-1}} \pdl\) .
& \VIId \cr}$$
%

The corresponding matrix integral
\eqn\VIIe{ \Z(\GL)=  {\int dM\, e^{{i^{p^2+1}\over 2(p+1)}
\tr \[ \(M+(-i)^{p+1} \GL)\)^{p+1}
\]_{>{\rm lin}}}
\over \int dM e^{-{1\over 4}\tr \sum_{k=0}^{p-1}M \GL^k M\GL^{p-1-k}}  }
}
may then be handled as in equations \IIm--\IIu. One considers the set of
functions $z^{(j)}$ defined by
\eqn\VIIea{ z^{(j)}(\Gl)= \Gl^{-j} D^j z(\Gl) , \qquad j\ge 0\ . }
{}From eq.~\VIIc, it follows that
$$D^{r (p-1)+j }z= \Gl^{r(p-1)+j} z^{(j)}\quad \mod ( D^0 z, \ldots,
D^{r(p-1)+j-1}z).$$ Thus the analogue of \IIu\ reads
\eqn\VIIeb{\Z(\GL)={\vert \Gl^0 z^{(0)},\ldots,\ \Gl^{p-1}z^{(p-1)},
\ \Gl^p z^{(0)}, \ldots \vert
\over \vert \Gl^0,\ \Gl^1,\ \Gl^2, \cdots, \Gl^{N-1} \vert}\ .}
The $p$ functions $z=z^{(0)}, \ldots, z^{(p-1)}$ have asymptotic expansions
\eqn\VIIec{z^{(j)}(\Gl) =\sum_0^{\infty}c_k^{(j)} \Gl^{-k(p+1)}\ ,  }
with coefficients $c_k ^{(j)}$; in the sequel the latter are
regarded as periodic in $j$
of period $p$: $c_k^{(j+p)}=c_k^{(j)}$.
One then proceeds as in sec.~2.2, introducing Schur functions,
with the result (analogous to \IIak) that
\eqn\VIIf{
\Z_k(\GL)=\sum_{n_0+\cdots+n_{N-1}=k} c_{n_0}^{(0)} c_{n_1}^{(1)} \cdots
c_{n_{N-1}}^{(N-1)} \left\vert\matrix{
p_{(p+1)n_0}& & \cr
&\ddots & \cr
&  &  p_{(p+1)n_{N-1}} \cr }\right\vert
}
is independent of $N$ for $N\ge (p+1)k$.

The steps of sec.~2.2 may then be followed sequentially to prove that
$Z_k$, computed now for $N=(p+1)k$, is independent of the $\Gth_{rp}=
\tr \GL^{-rp}$.  When differentiating $Z_k$ with respect to $\Gth_{rp}$,
the only non-trivially non vanishing terms are those for which the
derivative acts on one of the last $rp$ lines, where $r$ is
a multiple of $p+1$. The discussion of such a case then appeals to the
following identity generalizing \IIbb\
\eqn\VIIg{\det \BBL\Go ^{ij} z^{(j)}(\Go^i \Gl)\BBR_{0\le i,j \le p-1}
= {\rm constant} }
where $\Go$ is a $p$-th root of unity. This is proved by differentiating the
determinant, using the relations \VIIea\ between the functions $z^{(j)}$.
This implies a family of identities on the coefficients
\eqn\VIIh{\CC _{kp} \equiv\sum_{{{\scriptstyle \sum n_i = kp }\atop {
\scriptstyle n_i=i +\pi(i) \,\mod p}}}
\Ge_{\pi}\, c_{n_0}^{(0)}\ldots c_{n_{p-1}}^{(p-1)}=0 }
where the summation runs over the configurations of indices $n_i$
that may be written as indicated, with $\pi$ a permutation  of the $p$
integers $0, \ldots, p-1$.

On the other hand, as before,
the only contributions to $\dd {Z}{\Gth_{rp(p+1)}}$ come from Young
tableaux with a square-rule shape (as in eq.~\equerr), and one finds that
\eqn\VIIi{\dd {Z}{\Gth_{rp(p+1)}} = \sum_{l=0}^{rp-1} (-1)^l
\sum_{i=0}^{rp-1-l}\( 2 c_{rp-l}^{(i)}\GD_l^{(i+j+1)}
+\sum_{j=1}^{p-1} c_{rp-l}^{(i+j)}\GD_l^{(i+j+1)} \) \ ,}
where the $\GD$'s are determinants generalizing those of \IIbh:
\eqn\VIIj{\GD_s^{(j)}=
\left\vert\matrix{c_1^{(j)}&c_2^{(j+1)}&\ldots&c_s^{(j+s-1)}\cr
                           1&c_1^{(j+1)}&\ldots&c_{s-1}^{(j+s-1)}\cr
			   \vdots & \ddots& \ddots &\cr
			   0& \ldots & 1 & c_1^{(j+s-1)}\cr }\right\vert \ .}
The expression \VIIi\ may be recast in the form
\ifx\answ\bigans\eqn\VIIk
{\dd {Z}{\Gth_{rp(p+1)}} = \sum_{s=1}^{r}\sum_{i=0}^{p-1}
(-1)^{(s-1)p+i}
\BBL\sum_{j=0}^{p-1-i}\big((r-s+1)(p+1)-i\big) +
\sum_{j=p-i}^{p-1}\BBR c_{(r-s+1)p-i}^{(j)}\GD_{i+(s-1)p}^{(j+1)} \ .}
\else\eqnn\VIIk
$$\eqalignno{\dd {Z}{\Gth_{rp(p+1)}}& = \sum_{s=1}^{r}\sum_{i=0}^{p-1}
(-1)^{(s-1)p+i}
\BBL\sum_{j=0}^{p-1-i}\big((r-s+1)(p+1)-i\big) +
\sum_{j=p-i}^{p-1}\BBR
\cr &\qquad\qquad\qquad c_{(r-s+1)p-i}^{(j)}\GD_{i+(s-1)p}^{(j+1)} \ .
&\VIIk \cr}$$
\fi
It appears that the combination of $c$ and $\GD$ in the
summand in \VIIk, namely
$c_{sp-i}^{(j)}\GD_{i}^{(j+1)} $,
is, up to a sign, the coefficient of $c_{sp-i}^{(j)}$ in the constraint
$\CC_{sp}$ of eq. \VIIh
\eqn\VIIl{c_{sp-i}^{(j)}\GD_{i}^{(j+1)} =(-1)^{{(p-1)(p-2)\over 2}+i }
{\sum}' \Ge_{\pi} c_{n_0}^{(0)}\ldots c_{n_{p-1}}^{(p-1)} }
with the sum $\sum'$ subject to the same constraints as in \VIIh\ and
to $n_j=sp-i$.
Using this  fact and after some reshuffling, one finds that
$\pd Z/\pd \Gth_{rp(p+1)}$
is proportional to the constraint $\CC_{rp}$ and thus vanishes,
\eqn\VIIm{\dd {Z}{\Gth_{rp(p+1)}} =(-1)^{{(p-1)(p-2)\over 2}} r(p+1)
\CC_{rp} =0.}
The last part of the argument is carried out as in the end of sec.~2.2,
thus completing the proof of the independence of $Z$ with respect to the
$\Gth_{rp}$.

One then proceeds as in sec.~3, deriving the higher KdV hierarchies
associated with a differential operator $Q_p$ of order $p$ depending
on $p-1$ functions, and as in sec.~4, writing the generalized Airy
equation satisfied by $Z$. For example, in the case $p=3$, we have
\eqnn\VIIn
$$\eqalignno{\Big\{-\inv{8}t_j & +\CD_j^3 +\sum_{k, k\ne j} \[
\CD_j \inv{t_j-t_k}(\CD_j-\CD_k)+ \inv{t_j-t_k}(\CD_j^2-\CD_k^2) \]\cr
&+ \sum_{{{\scriptstyle k,l}
\atop {\scriptstyle j\ne k \ne l\ne k} }} \( \inv{(t_j-t_k)(t_j-t_l)}\CD_j
+{\rm circ.}\)
\Big\} Z =0 & \VIIn\cr}$$
with the following notations
\eqnn\VIIo
$$\eqalignno{t_j&= \Gl_j^3 \cr
\pd_j&= \dd{}{t_j}\cr
\CD_j&= e^{-\frac{3}{8}\GS_k \Gl_k^4}\prod_{k,l}\(\Gl_k^2+\Gl_k\Gl_l
+\Gl_l^2\)^ {\oh} \pd_j
 e^{\frac{3}{8}\GS_k \Gl_k^4}\prod_{k,l}\(\Gl_k^2+\Gl_k\Gl_l
+\Gl_l^2\)^ {-\oh}  \cr
&= \pd_j +a_j & \VIIo \cr
a_j=& \oh \Gl_j -\inv{3\Gl_j^2}  \sum_k {2\Gl_j+\Gl_k\over \Gl_j^2+
\Gl_k\Gl_j+\Gl_k^2} \cr
}$$
{}From this system of equations, the strong reader will be able to
extract the expression of the generators of
the $W_3$ algebra, in a way similar to
\IVh, and to calculate the analogues of the genus expansion and of
the singular behavior discussed in sec.~5 and 6 \dots

\vskip2truecm
{\bf Acknowledgements.}\par
\noindent It is a pleasure to acknowledge some inspiring
correspondence
with M. Kontsevich and to thank
M. Bauer for his assistance in algebraic calculations as well as in the
elaboration of Lemma 5 and P. Ginsparg for a critical reading
of the manuscript.

\listrefs
\end